\documentclass[aps,onecolumn]{revtex4}
\usepackage{amsfonts}
\usepackage{amssymb,epsf}

\begin{document}

\title{$F(R)$ gravity's rainbow and its Einstein counterpart}
\author{S. H. Hendi$^{1,2}$\thanks{
email address: hendi@shirazu.ac.ir}, B. Eslam Panah$^{1}$ \thanks{
email address: behzad.eslampanah@gmail.com}, S. Panahiyan$^{1,3}$ \thanks{
email address: sh.panahiyan@gmail.com}, and M. Momennia$^{1}$ \thanks{
email address: momennia1988@gmail.com}}
\affiliation{$^1$ Physics Department and Biruni Observatory, College of Sciences, Shiraz
University, Shiraz 71454, Iran\\
$^{2}$ Research Institute for Astronomy and Astrophysics of Maragha (RIAAM),
Maragha, Iran\\
$^3$ Physics Department, Shahid Beheshti University, Tehran 19839, Iran}

\begin{abstract}
Motivated by UV completion of general relativity with a modification of a
geometry at high energy scale, it is expected to have an energy dependent
geometry. In this paper, we introduce charged black hole solutions with
power Maxwell invariant source in the context of gravity's rainbow. In
addition, we investigate two classes of $F(R)$ gravity's rainbow solutions.
At first, we study energy dependent $F(R)$ gravity without energy momentum
tensor, and then we obtain $F(R)$ gravity's rainbow in the presence of
conformally invariant Maxwell source. We study geometrical properties of the
mentioned solutions and compare their results. We also give some related
comments regarding to thermodynamical behavior of the obtained solutions and
discuss thermal stability of the solutions.
\end{abstract}

\maketitle

\section{Introduction}

An accelerated expansion of the Universe was confirmed by various
observational evidences. The luminosity distance of Supernovae type Ia \cite%
{Perlmutter1,Perlmutter2}, the anisotropy of cosmic microwave background
radiation \cite{CMBR}, and also wide surveys on galaxies \cite{Gal} confirm
such accelerated expansion. On the other hand, baryon oscillations \cite{BO}%
, large scale structure formation \cite{LSS}, and weak lensing \cite{WL}
also propose such an accelerated expansion of the Universe.

After discovery of such an expansion in $1998$, understanding its
theoretical reasons presents one of the fundamental open questions in
physics. Identifying the cause of this late time acceleration is a
challenging problem in cosmology. Physicists are interested in considering
this accelerated expansion in a gravitational background and they proposed
some candidates to explain it. For example, a positive cosmological constant
leads to an accelerated expansion, but it is plagued by the fine tuning
problem \cite{Sahni1,Sahni2,Sahni4,Sahni5,Sahni6}. In other words, the left
hand side of Einstein equations can modify by the cosmological constant as a
geometrical modification or it can be interpreted as a kinematic term on the
right hand side with the equation of state parameter $w=-1$. By considering $%
w<-1/3$ for a source term, it is possible to further modify this approach.
This consideration has interpretation of Dark Energy which has been
investigated in literature \cite%
{Armendariz1,Armendariz2,Armendariz3,Armendariz5,Armendariz6,Armendariz9,Armendariz10}%
. In dark energy models, the acceleration expansion of the universe is due
to an unknown ingredient added to the cosmic pie. The effects of this
unknown ingredient is extracted by modifying the stress energy tensor of the
Einstein equation with a matter which is different from than the usual
matter and radiation components.

On the other hand, it is proposed that the presence of accelerated expansion
of the universe indicate that the standard general relativity requires
modification. To do so, one can generalize the Einstein field equations to
obtain a modified version of gravity. There are different branches of
modified gravity with various motivations, such as brane world cosmology
\cite{Gergely1,Gergely2,Gergely3}, Lovelock gravity \cite%
{LovelockA,LovelockB,Lovelock2a,Lovelock2d,Lovelock2f}, scalar-tensor
theories \cite%
{JordanA,JordanB,JordanC,JordanD,JordanE,Jordan22,Jordan23,Jordan24}, and $%
F(R)$ gravity \cite%
{FR1,FR11,Akbar,FR2,FR3,FR4,FR5,FR7,FR8,FR9,FR99,FR10,FR15,FR16,FR166,HendiGRG}%
.

Modifying general relativity opens a new way to a large class of alternative
theories of gravity ranging from higher dimensional physics \cite%
{Dvali1,Dvali2,Dvali3} to non-minimally coupled (scalar) fields \cite%
{Caresia1,Caresia2,Caresia3,Caresia4}. Regarding various models of modified
gravity, we will be interested in $F(R)$ gravity \cite%
{Capozziello1,Capozziello2,Capozziello3,Capozziello4,Capozziello5,Capozziello6,Capozziello7}
based on replacing the scalar curvature $R$ with a generic analytic function
$F(R)$ in the Hilbert-Einstein action. Some viable functional forms of $F(R)$
gravity may be reconstructed starting from data and physically motivated
issues.

However, the field equations of $F(R)$ gravity are complicated fourth order
differential equations, and it is not easy to find exact solutions. In
addition, adding a matter field to $F(R)$ gravity makes the field equations
much more complicated. On the other hand, regarding constant curvature
scalar model as a subclass of general $F(R)$ gravity can simplify the field
equations. Also, one can extract exact solutions of $F(R)$ gravity coupled
to a traceless energy momentum tensor with constant curvature scalar \cite%
{Moon}. For example, considering exact solutions of $F(R)$ gravity with
conformally invariant Maxwell (CIM) field as a matter source has been
investigated \cite{Sheykhi1,HendiES}.

General relativity coupled to a nonlinear electrodynamics attracts
significant attentions because of its specific properties in gauge/gravity
coupling. Interesting properties of various models of the nonlinear
electrodynamics have been investigated before \cite%
{NLED1,NLED2,NLED3,NLED6,NLED8,NLED9,NLED10,NLED11,NLED12,NLED13,NLED14,NLED15,NLED16}%
. Power Maxwell invariant (PMI) theory is one of the interesting branches of
the nonlinear electrodynamics which its Lagrangian is an arbitrary power of
Maxwell Lagrangian \cite{Hassaine2008A,Hassaine2008B,Hassaine2008D}. The PMI
theory is more interesting with regard to Maxwell field, and for unit power,
it reduces to linear Maxwell theory. This nonlinear electrodynamics enjoys
conformal invariancy when the power of Maxwell invariant is a quarter of
spacetime dimensions, and this is one of the attractive properties of this
theory. In other words, one can obtain traceless energy-momentum tensor for
a special case "$power=dimensions/4$" which leads to conformal invariancy.
It is notable to mention that this idea has been considered to take
advantage of the conformal symmetry to construct the analogues of the four
dimensional Reissner-Nordstr\"{o}m solutions with an inverse square electric
field in arbitrary dimensions \cite{Hassaine2007A}.

From the gravitational point of view, it is possible to show that the
electric charge and cosmological constant can be extracted, simultaneously,
from pure $F(R)$ gravity (without matter field: $T_{\mu \nu }=0$) \cite%
{HendiGRG}. In this paper, we are going to obtain $d$-dimensional charged
black hole solutions from gravity's rainbow, pure $F(R)$ gravity's rainbow
as well as $F(R)$ gravity's rainbow with CIM source and compare them to
obtain their direct relation.

In order to build up special relativity from Galilean theory, one has to
take into account an upper limit for velocity of particles. The same method
could be used to restrict particles from obtaining energies no more than
specific energy, the so-called Planck energy scale. This upper bound of
energy may modify dispersion relation which is known as double special
relativity \cite{double}. Generalization of this doubly special relativity
to incorporate curvature leads to gravity's rainbow \cite{rain}. In
gravity's rainbow, spacetime is a combination of the temporal and spatial
coordinates as well as energy functions. The existence of such energy
functions indicates that, the particle probing the spacetime can acquire
specific range of energies which in essence leads to formation of a rainbow
of energy.

There are several features for gravity's rainbow which highlight the
importance of such generalization. Among them one can point modification in
energy-momentum dispersion relation which is supported by studies that are
conducted in string theory \cite{st}, loop quantum gravity \cite{loop} and
experimental observation \cite{obs}. Also, existence of remnant for black
holes \cite{rem} which is proposed to be a candidate for solving the
information paradox \cite{paradox}. In addition, this theory admits the
usual uncertainty principle \cite{uncertain1,uncertain2}.

Recently, there has been a growing interest in energy dependent spacetimes
\cite{rainbow1,rainbow3,rainbow5,rainbow6,rainbow7,rainbow9}. Different
classes of black holes have been investigated in the context of gravity's
rainbow \cite{blackRainbow1,blackRainbow2,blackRainbow3,blackRainbow4}. The
hydrostatic equilibrium equation of stars in the presence of gravity's
rainbow has been obtained \cite{TOV}. Furthermore, a study regarding the
gravity's rainbow and compact stars has been done \cite{compact rainbow}. In
Ref. \cite{wormhole}, the effects of gravity's rainbow for wormholes have
been investigated. Moreover, the influences of gravity's rainbow on
gravitational force have been investigated \cite{force rainbow}. Also,
Starobinsky model of $F(R)$ theory in gravity's rainbow has been studied in
Ref. \cite{Starobinsky rainbow}. In addition, gravity's rainbow has
interesting effects on the early universe \cite%
{Nonsingular1,Nonsingular2,Nonsingular3}.

The main motivations for studying black holes in the presence of gravity's
rainbow given as follows. First of all, due to the high energy properties of
the black holes, it is necessary to consider quantum corrections of
classical perspectives. One of the methods to include quantum corrections of
gravitational fields is by considering an energy dependent spacetime. In
fact, it was shown that the quantum correction of gravitational systems
could be observed in dependency of spacetime on the energy of particles
probing it which is gravity's rainbow point of view \cite%
{uncertain1,Smolin1,Grattini1}. Since we are modifying our point of view to
an energy dependent spacetime, it is expected to find its effects on the
properties of black holes, especially in the context of their
thermodynamics. This is another motivation for considering gravity's rainbow
generalization. Also, there are specific achievements for gravity's rainbow
in the context of black holes which among them one can name: modified
uncertainty principle \cite{uncertain1,uncertain2}, existence of remnants
for the black holes \cite{rem,blackRainbow2}, furnishing a bridge towards
Horava---Lifshitz gravity \cite{Emanuel}, providing possible solution toward
information paradox \cite{rem} and finally being UV completion of Einstein
gravity \cite{Mag}. In addition, as it was pointed out before, in the
context of cosmology, it presents a possible solution toward big bang
singularity problem \cite{Nonsingular1,Nonsingular2,Nonsingular3}. On the
other hand, $F(R)$ gravity provides correction toward gravitational sector
of the Einstein theory of gravity. The importance of this correction is
highlighted in the context of black holes. In order to have better picture
regarding the physical nature of the black holes, one may consider high
energy regime effects, as well (considering that the concepts of Hawking
radiation was derived by studying black holes in semi classical/quantum
regime). Here, we apply $F(R)$ gravity's rainbow to find the effects of $%
F(R) $ generalization as well as energy dependency of spacetime on the black
hole solutions.

The outline of our paper is as follows. In Section II, we are going to
investigate black hole solutions in Einstein-gravity's rainbow with PMI and
CIM fields. Then, we want to investigate conserved and thermodynamic
quantities of the solutions and check the first law of thermodynamics. In
Section III, we will obtain black hole solutions of $F(R)$ gravity's rainbow
with CIM source and check the first law of thermodynamics. We also discuss
thermal stability of these solutions and criteria governing
stability/instability in Sec. \ref{Stability}. Then, we consider pure $F(R)$
gravity's rainbow and compare these solutions with $F(R)$ gravity's rainbow
with CIM source and give some related comments regarding to its
thermodynamical behavior. Finally, we finish our paper by some conclusions.


\section{Einstein Gravity's Rainbow in the Presence PMI Field}

Here, we are going to introduce $d$-dimensional solutions of the
Einstein-gravity's rainbow in the presence of PMI field with the following
Lagrangian
\begin{equation}
\mathcal{L}=R-2\Lambda +(\kappa \mathcal{F})^{s},  \label{ActionEN-PM}
\end{equation}%
where $R$ and $\Lambda $ are, respectively, the Ricci scalar and the
cosmological constant. In Eq. (\ref{ActionEN-PM}), the Maxwell invariant is $%
\mathcal{F}=F_{\mu \nu }F^{\mu \nu }$ where $F_{\mu \nu }=\partial _{\mu
}A_{\nu }-\partial _{\nu }A_{\mu }$ is the electromagnetic tensor field and $%
A_{\mu }$ is the gauge potential. As we mentioned before, in the limit $s=1$
with $\kappa =-1,$the Lagrangian (\ref{ActionEN-PM}) reduces to the
Lagrangian of Einstein-Maxwell gravity. Since the Maxwell invariant is
negative, henceforth we set $\kappa =-1$, without loss of generality. Using
the variation principle, we can find the field equations the same as those
obtained in Ref. \cite{HendiPMIepjc}.

\subsection{Black hole Solutions}

Here, we will obtain charged rainbow black hole solutions with negative
cosmological constant in $d$-dimensions. It is notable that the charged
rainbow black hole solutions in Einstein gravity coupled to nonlinear
electromagnetic fields have been studied in Ref. \cite{HendiPEMrainbow}. In
this paper, we want to extend the spacetime to $d$-dimensions and obtain
black hole solutions in the presence of PMI field as a matter source. The
rainbow metric for spherical symmetric spacetime in $d$-dimensions can be
written as%
\begin{equation}
ds^{2}=-\frac{\psi \left( r\right) }{f\left( E\right) ^{2}}dt^{2}+\frac{1}{%
g\left( E\right) ^{2}}\left[ \frac{dr^{2}}{\psi \left( r\right) }%
+r^{2}d\Omega ^{2}\right] ,  \label{metric}
\end{equation}%
where%
\begin{equation}
d\Omega ^{2}=d\theta _{1}+\sum_{i=2}^{d-2}\prod\limits_{j=1}^{i-1}\sin
^{2}\theta _{j}d\theta _{i}^{2}.
\end{equation}

Using the metric (\ref{metric}) with the field equations, one can obtain the
following solutions for the metric function as well as gauge potential
\begin{equation}
\psi \left( r\right) =1-\frac{m}{r^{d-3}}-\frac{2\Lambda r^{2}}{\left(
d-1\right) \left( d-2\right) g\left( E\right) ^{2}}+\left\{
\begin{array}{cc}
\frac{-2^{\frac{d-1}{2}g\left( E\right) ^{d-3}\left( qf(E)\right) ^{d-1}}}{%
r^{d-3}}\ln (\frac{r}{l}), & s=\frac{d-1}{2} \\
&  \\
\frac{r^{2}\left( 2s-1\right) ^{2}\left( \frac{2\left[ qf\left( E\right)
g\left( E\right) \left( 2s-d+1\right) \right] ^{2}r^{\frac{-2d+4}{2s-1}}}{%
\left( 2s-1\right) ^{2}}\right) ^{s}}{\left[ \left( d-1\right) \left(
d-2\right) -2\left( d-2\right) s\right] g\left( E\right) ^{2}}, & otherwise%
\end{array}%
\right. ,  \label{g(r)BTZ}
\end{equation}%
\begin{equation}
A_{\mu }=h(r)\delta _{\mu }^{t},  \label{gauge Pot}
\end{equation}%
where the consistent $h(r)$ function is $-q\ln \frac{r}{l}$ or $-qr^{\frac{%
2s-d+1}{2s-1}}$ for $s=\frac{d-1}{2}$ and $s\neq \frac{d-1}{2}$,
respectively. In addition, $m$ and $q$ are integration constants which are,
respectively, related to the mass and electric charge of the black hole. We
should also mention that we consider $s>1/2$ for obtaining well-behaved
electromagnetic field. It is notable that by replacing $s=1$ and $g(E)=f(E)=1
$, the solutions (\ref{g(r)BTZ}) reduce to the following higher dimensional
Reissner-Nordstr\"{o}m black hole solutions%
\begin{equation}
\psi \left( r\right) =1-\frac{m}{r^{d-3}}-\frac{2\Lambda r^{2}}{\left(
d-1\right) \left( d-2\right) }+\frac{2\left( d-3\right) q^{2}}{\left(
d-2\right) r^{2\left( d-3\right) }}.
\end{equation}

\begin{figure}[tbp]
$%
\begin{array}{ccc}
\epsfxsize=5.5cm \epsffile{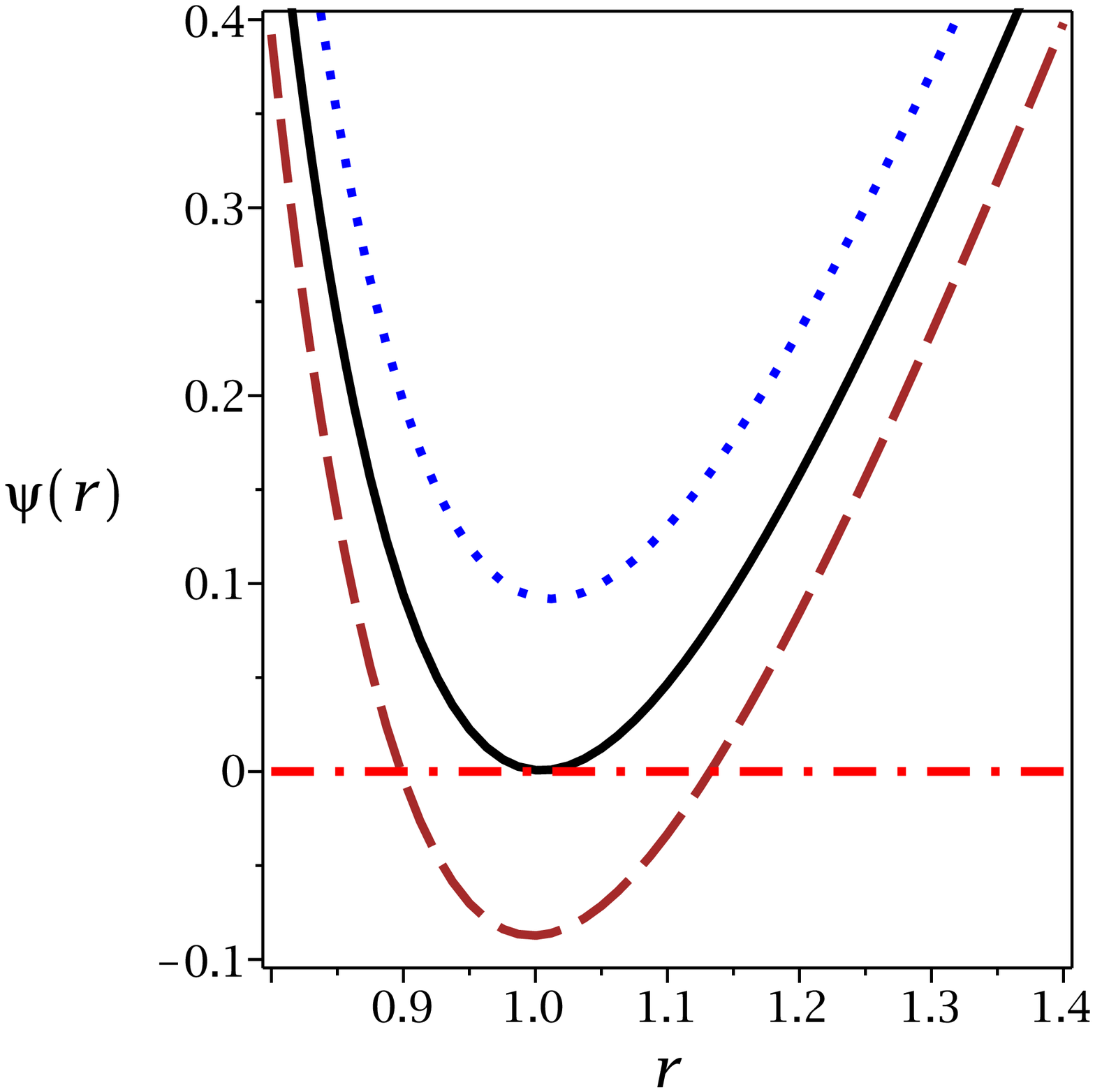} & \epsfxsize=5.5cm %
\epsffile{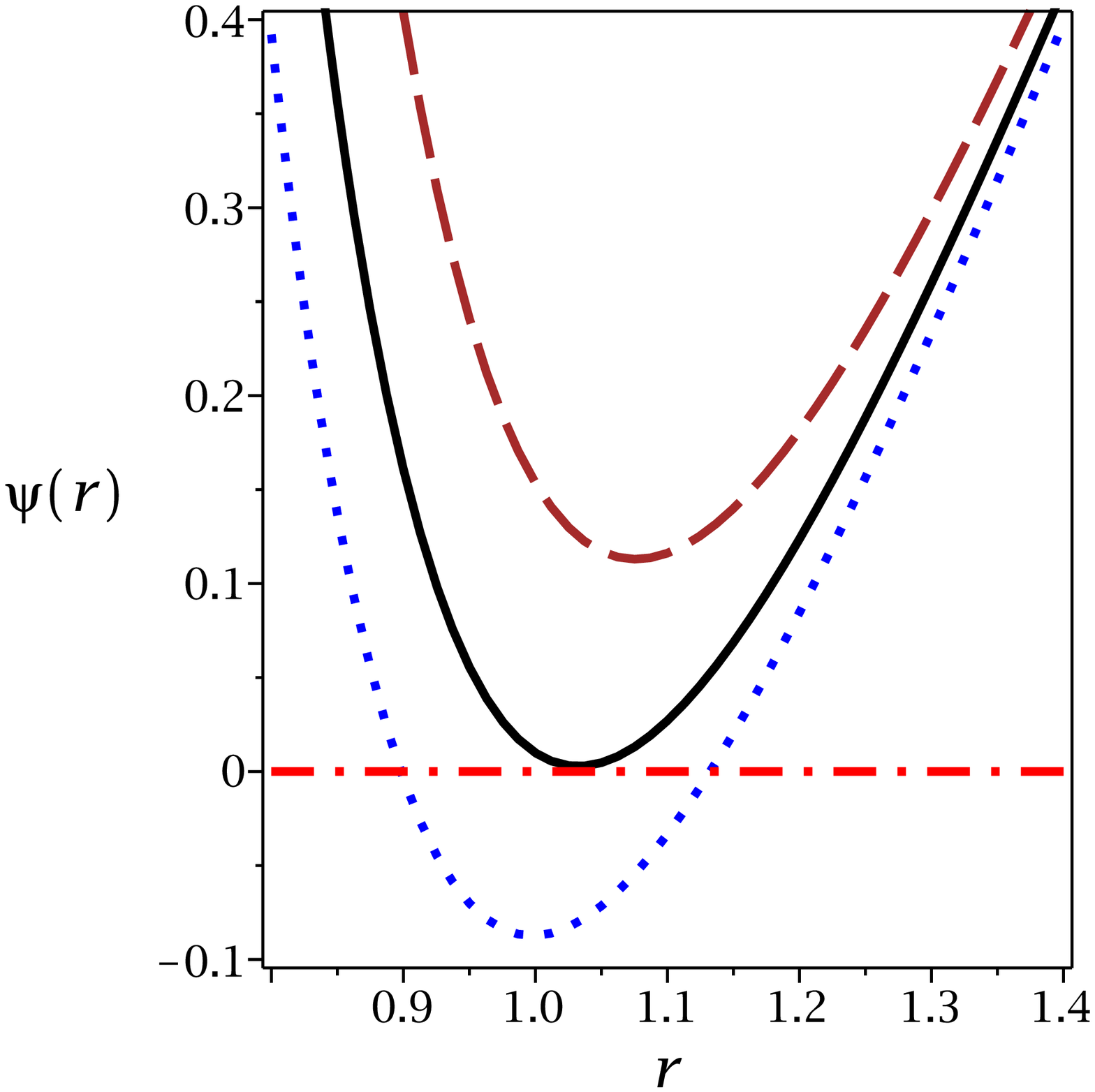} & \epsfxsize=5.5cm \epsffile{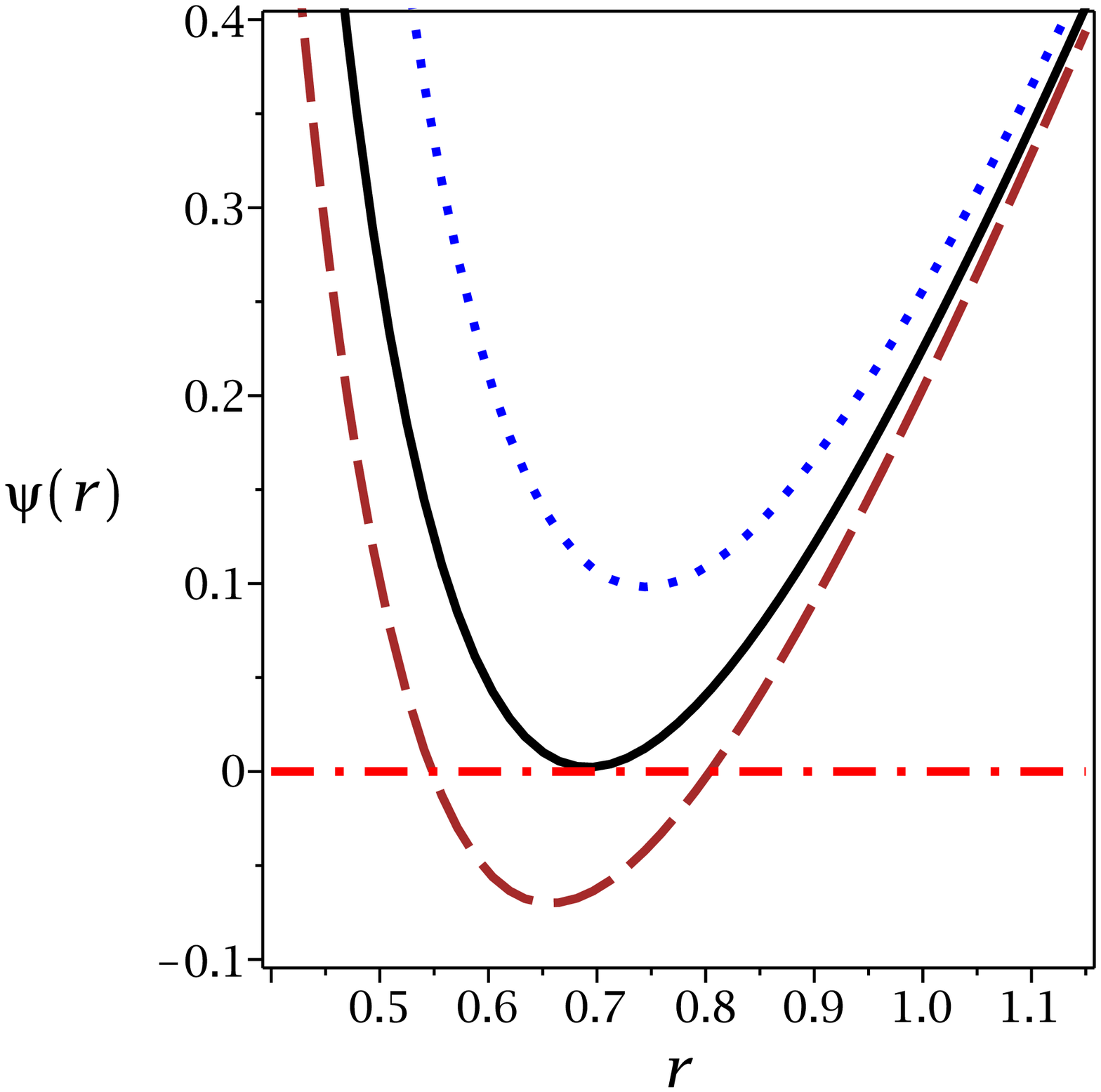}%
\end{array}
$%
\caption{$\protect\psi (r)$ versus $r$ for $g(E)=1.1$, $f(E)=1.1$, $\Lambda
=-1$ and $d=4$.\newline
Left diagram: for $s=0.7$, $q=1$, $m=1.95$ (doted line), $m=1.87$
(continuous line) and $m=1.80$ (dashed line).\newline
Middle diagram: for $s=0.7$, $m=2$, $q=0.5$ (doted line), $q=0.7$
(continuous line) and $q=1.1$ (dashed line).\newline
Right diagram: for $m=2$, $q=1$, $s=1.125$ (doted line), $s=1.108$
(continuous line) and $s=1.095$ (dashed line).}
\label{Fig1}
\end{figure}

\begin{figure}[tbp]
$%
\begin{array}{ccc}
\epsfxsize=7cm \epsffile{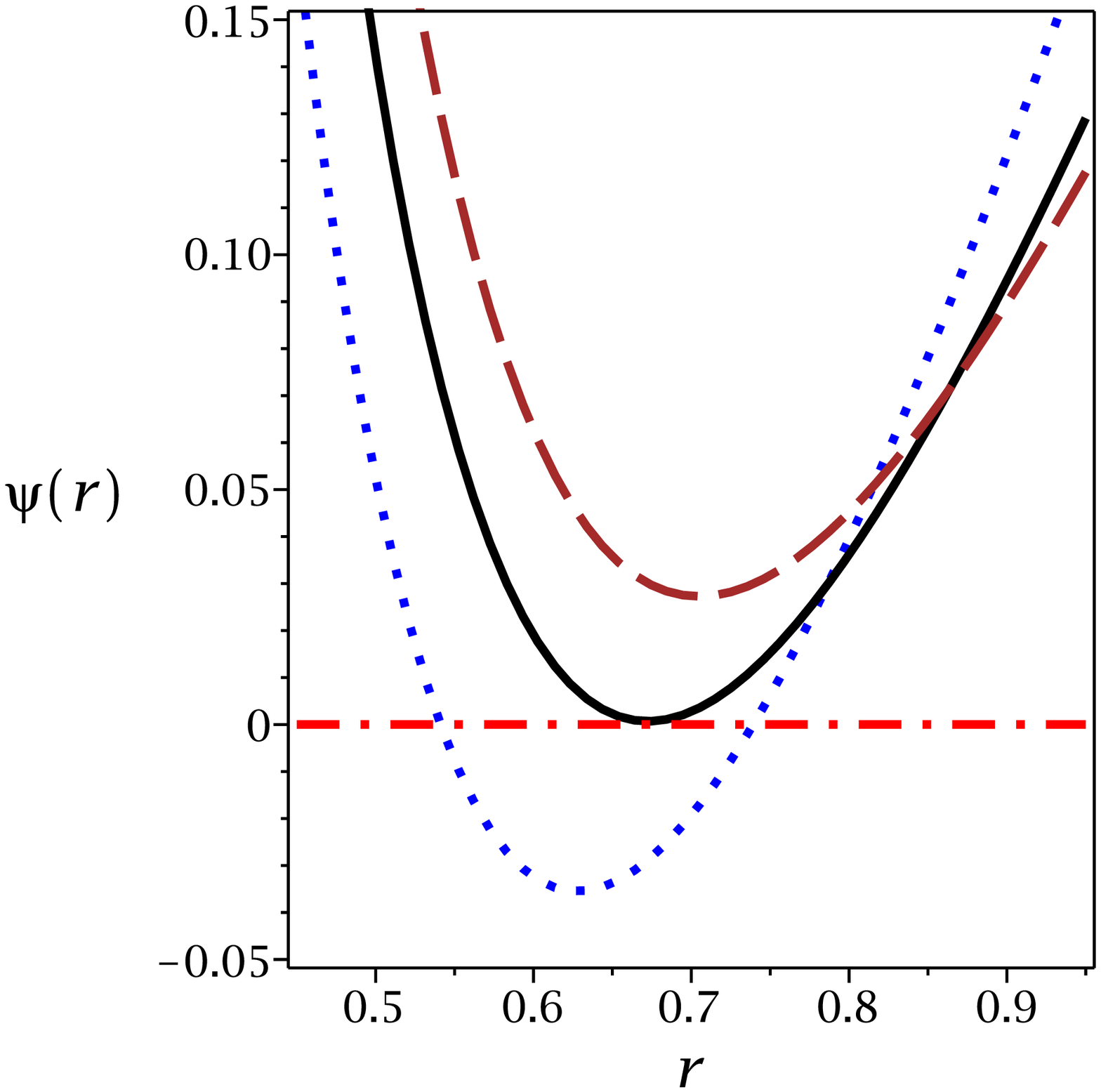} & \epsfxsize=7cm %
\epsffile{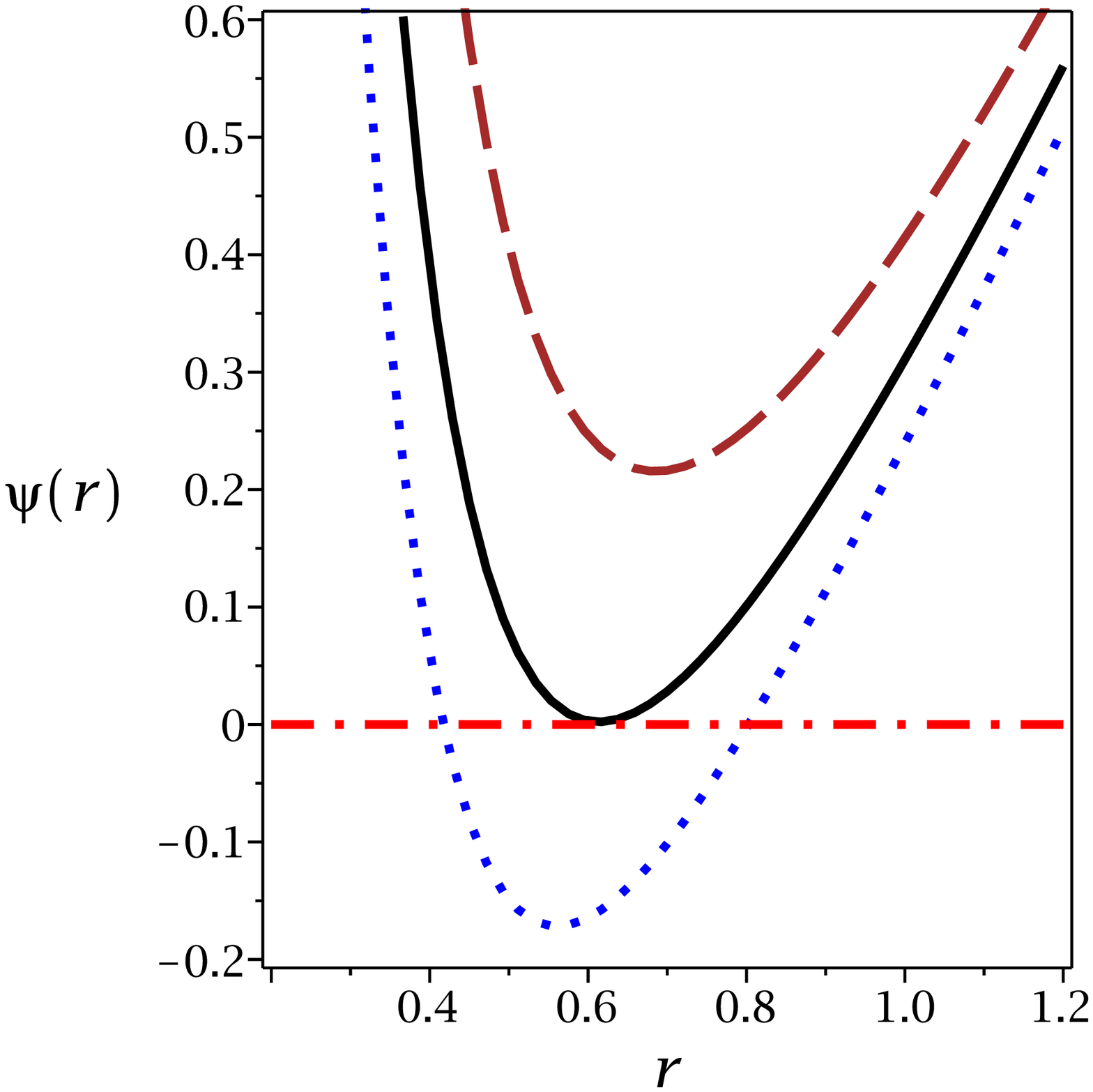} &
\end{array}
$%
\caption{$\protect\psi (r)$ versus $r$ for $m=2$, $q=0.8$, $\Lambda =-1$,
and $d=4$.\newline
Left diagram: for $f(E)=1.00$, $s=1.06$, $g(E)=1.40$ (dotted line), $%
g(E)=0.86$ (continuous line) and $g(E)=0.68$ (dashed line). \newline
Right diagram: for $g(E)=1.0$, $s=1.1$, $f(E)=0.95$ (dotted line), $%
f(E)=1.02 $ (continuous line) and $f(E)=1.10$ (dashed line).}
\label{Fig2}
\end{figure}

In order to investigate the geometrical structure of these solutions, we
first look for the essential singularity(ies). The Ricci scalar can be
written as%
\begin{equation}
R=\frac{2d}{d-2}\Lambda +\left\{
\begin{array}{cc}
\frac{2^{\frac{d-1}{2}\left( qf(E)g(E)\right) ^{d-1}}}{r^{d-1}}, & s=\frac{%
d-1}{2} \\
\frac{4s-d}{d-2}\left( \frac{2\left[ qf\left( E\right) g\left( E\right)
\left( 2s-d+1\right) \right] ^{2}r^{\frac{-2d+4}{2s-1}}}{\left( 2s-1\right)
^{2}}\right) ^{s}, & otherwise%
\end{array}%
\right. ,  \label{RiccichrBTZ}
\end{equation}%
and the behavior of Kretschmann scalar is%
\begin{eqnarray}
\lim_{r\longrightarrow 0}R_{\alpha \beta \gamma \delta }R^{\alpha \beta
\gamma \delta } &\longrightarrow &\infty ,  \nonumber \\
\lim_{r\longrightarrow \infty }R_{\alpha \beta \gamma \delta }R^{\alpha
\beta \gamma \delta } &=&\frac{8d}{\left( d-1\right) \left( d-2\right) ^{2}}%
\Lambda ^{2},
\end{eqnarray}%
therefore, confirm that there is a curvature singularity at $r=0$. In
addition, the Kretschmann scalar is $\frac{8d}{\left( d-1\right) \left(
d-2\right) ^{2}}\Lambda ^{2}$ for $r\longrightarrow \infty $, which confirms
that the asymptotical behavior of the charged rainbow black hole is adS. It
is worthwhile to mention that the asymptotical behavior of these solutions
is independent of the rainbow functions and the power of PMI source. This
independency comes from the fact that rainbow functions are sensible in high
energy regime such as near horizon and one expects to ignore its effects far
from the black hole (we only consider high energy regime). In addition, at
large distances, the electric field of Maxwell and PMI theories vanishes,
and therefore, one may expect to ignore the effects of electric charge far
from the origin. In order to investigate the possibility of the horizon, we
plot the metric function versus $r$ in Figs. \ref{Fig1} and \ref{Fig2}. It
is evident that depending on the choices of values for different parameters,
we may encounter with two horizons (inner and outer horizons), one extreme
horizon and without horizon (naked singularity).


\textbf{Special Case }$s=\frac{d}{4}$\textbf{:}

Here, we are going to investigate the special case $s=\frac{d}{4}$, the
so-called conformally invariant Maxwell field (for more details, see \cite%
{HendiES,Hassaine2007A}). It was shown that for $s=\frac{d}{4}$, the
energy-momentum tensor will be traceless and the corresponding electric
field will be proportional to $r^{-2}$ in arbitrary dimensions, as it takes
place for the Maxwell field in $4$-dimensions. Therefore, we consider $s=%
\frac{d}{4}$ into Eqs. (\ref{g(r)BTZ}) and (\ref{gauge Pot}) to obtain%
\begin{equation}
\psi \left( r\right) =1-\frac{m}{r^{d-3}}-\frac{2\Lambda r^{2}}{\left(
d-1\right) \left( d-2\right) g\left( E\right) ^{2}}+\frac{2^{\frac{d-4}{4}%
}\left( qf\left( E\right) g(E)\right) ^{\frac{d}{2}}}{g(E)^{2}r^{d-2}},
\label{g(r)CIM}
\end{equation}%
\begin{equation}
h(r)=-\frac{q}{r}.  \label{h(r)CIM}
\end{equation}%
\ \ \

\begin{figure}[tbp]
$%
\begin{array}{c}
\epsfxsize=7cm \epsffile{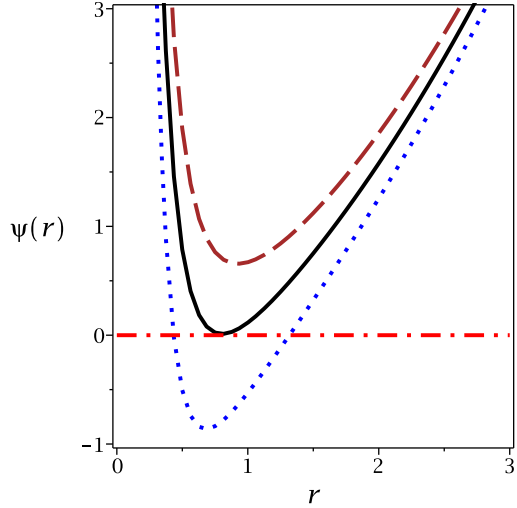}%
\end{array}
$%
\caption{$\protect\psi (r)$ versus $r$ for $q=1$, $d=4$, $f(E)=1.1$, $%
\Lambda =-1$, $g(E)=1.0$, $m=3.00$ (dotted line), $m=2.46$ (continuous line)
and $m=1.80$ (dashed line).}
\label{Fig3}
\end{figure}

Considering $g(E)=f(E)=1$ in the above equation, we obtain the higher
dimensional Reissner-Nordstr\"{o}m black hole solution. In order to make
more investigations regarding the properties of these solutions, we plot the
metric function (\ref{g(r)CIM}) in Fig. \ref{Fig3}. It is evident that the
behavior of Einstein-CIM-rainbow black holes (Eq. (\ref{g(r)CIM})) is
similar to Einstein-PMI-rainbow black holes (see Figs. \ref{Fig1} and \ref%
{Fig2}).

\subsection{Thermodynamics}

Here, we are going to calculate the conserved and thermodynamic quantities
and check the first law of thermodynamics for such black hole. In order to
calculate thermodynamic quantities, we start with temperature. Using the
definition of surface gravity, one can calculate the Hawking temperature of
the black hole as
\begin{equation}
T_{+}=\frac{1}{2\pi }\sqrt{-\frac{1}{2}\left( \nabla _{\mu }\chi _{\nu
}\right) \left( \nabla ^{\mu }\chi ^{\nu }\right) },  \label{Tem1PMI2}
\end{equation}%
where $\chi =\partial /\partial t$ is the Killing vector. So, the Hawking
temperature for the rainbow black holes in the presence of PMI source can be
written as
\begin{equation}
T_{+}=-\frac{1}{2\pi f(E)}\left\{ \frac{\Lambda r_{+}}{(d-2)g(E)}-\frac{%
(d-3)g(E)}{2r_{+}}\right\} -\mathcal{T},  \label{TemPMI2}
\end{equation}%
where $\mathcal{T}$ is
\begin{equation}
\mathcal{T}=\left\{
\begin{array}{cc}
\frac{2^{(d-5)/2}[qf(E)g(E)]^{d}}{\pi qf(E)^{2}g(E)^{2}r_{+}^{d-2}} & s=%
\frac{d-1}{2} \\
&  \\
\frac{2^{s-2}(2s-1)r_{+}}{(d-2)\pi f(E)g(E)}\left[ \frac{qf(E)g(E)(d-2s-1)}{%
(2s-1)r_{+}^{(d-2)/(2s-1)}}\right] ^{2s} & otherwise%
\end{array}%
\right. ,  \nonumber
\end{equation}%
in which $r_{+}$ satisfies $f(r=r_{+})=0$. Moreover, the electric potential $%
U$ is defined by \cite{Cvetic1,Cvetic2}
\begin{equation}
U=A_{\mu }\chi ^{\mu }\left\vert _{r\rightarrow reference }\right. -A_{\mu
}\chi ^{\mu }\left\vert _{r=r_{+}}\right.
\end{equation}

So for these black holes, we obtain
\begin{equation}
U=\left\{
\begin{array}{cc}
q\ln \left( \frac{r_{+}}{l}\right) & s=\frac{d-1}{2} \\
&  \\
qr_{+}^{\frac{(2s-3)}{(2s-1)}} & otherwise%
\end{array}%
\right. .  \label{ElecPoPMI1}
\end{equation}

The entropy of the black holes satisfies the so-called area law of entropy
in Einstein gravity. It means that the black hole's entropy equals to
one-quarter of horizon area \cite{Beckenstein1,Beckenstein2,Beckenstein3}.
Therefore, the entropy of the black holes in $d$-dimensions is
\begin{equation}
S=\frac{1}{4}\left( \frac{r_{+}}{g(E)}\right) ^{d-2}.  \label{entropyPMI}
\end{equation}

In order to obtain the electric charge of the black holes, one can calculate
the flux of the electromagnetic field at infinity, so we have
\begin{equation}
Q=\left\{
\begin{array}{cc}
-\frac{1}{\pi }2^{\frac{(d-3)}{2}}(d-1)\left[ qf(E)\right] ^{d-2} & s=\frac{%
d-1}{2} \\
&  \\
\frac{s(2s-1)}{8(2s-d+1)\pi qf(E)g(E)^{d-1}}\left[ \frac{\sqrt{2}%
(2s-d+1)qf(E)g(E)}{(2s-1)}\right] ^{2s} & otherwise%
\end{array}%
\right. .  \label{QPMI}
\end{equation}

The spacetime introduced in Eq. (\ref{metric}), have boundaries with
timelike $\left( \xi =\partial /\partial t\right) $\ Killing vector field.
It is straightforward to show that the total finite mass can be written as%
\begin{equation}
M=\frac{(d-2)m}{16\pi f(E)g(E)^{d-3}}.  \label{massPMI}
\end{equation}

Now, we are in a position to check the validity of the first law of
thermodynamics. To do so, one can employ the following relation
\begin{equation}
dM(S,Q)=\left( \frac{\partial M(S,Q)}{\partial S}\right) _{Q}dS+\left( \frac{%
\partial M(S,Q)}{\partial Q}\right) _{S}dQ.
\end{equation}

It is a matter of calculation to show that following equalities hold%
\begin{equation}
T=\left( \frac{\partial M}{\partial S}\right) _{Q}\ \ \ \ ,\ \ \ \ U=\left(
\frac{\partial M}{\partial Q}\right) _{S},  \label{TU}
\end{equation}%
which confirm that the first law of thermodynamics is valid for the obtained
thermodynamic and conserved quantities. We will address thermodynamic
stability in section 4.

\section{$F(R)$ Gravity's Rainbow in the Presence of CIM Field}

Here, we consider $F(R)=R+f(R)$ gravity's rainbow with the CIM field as a
matter source, which leads to a traceless energy-momentum tensor. For $d$%
-dimensions, the equations of motion for the $F(R)$ gravity's rainbow with
CIM source are
\begin{equation}
R_{\mu \nu }\left( 1+f_{R}\right) -\frac{g_{\mu \nu }}{2}F(R)+\left( g_{\mu
\nu }\nabla ^{2}-\nabla _{\mu }\nabla _{\nu }\right) f_{R}=8\pi \mathrm{T}%
_{\mu \nu },  \label{EqF(R)1}
\end{equation}%
\begin{equation}
\partial _{\mu }\left( \sqrt{-g}F^{\mu \nu }\mathcal{F}^{\left( \frac{d}{4}%
-1\right) }\right) =0,  \label{EqF(R)2}
\end{equation}%
in which $f_{R}=\frac{df(R)}{dR}$. In order to extract black hole solutions
in $F(R)$ gravity's rainbow coupled to the matter field, it is essential
that we consider a traceless energy-momentum tensor.

\subsection{Black hole Solutions}

We want to obtain the black hole solutions for constant scalar curvature ($%
R=R_{0}$= const). Using Eqs. (\ref{gauge Pot}) and (\ref{EqF(R)2}) with
metric (\ref{metric}), one can find
\begin{equation}
h(r)=\frac{-\mathcal{B}}{r},\ \ \ \&\ \ \ \ F_{tr}=\frac{\mathcal{B}}{r^{2}},
\end{equation}
where $\mathcal{B}$ is an integration constant.

Regarding the trace of Eq. (\ref{EqF(R)1}), one finds $R=R_{0}=\frac{%
df(R_{0})}{2\left( 1+f_{R}\right) -d}\equiv 4\Lambda $. Substituting the
mentioned $R_{0}$ into Eq. (\ref{EqF(R)1}), we obtain the following equation
\begin{equation}
R_{\mu \nu }\left( 1+f_{R}\right) -\frac{1}{d}g_{\mu \nu }R_{0}\left(
1+f_{R}\right) =8\pi \mathrm{T}_{\mu \nu }.  \label{Filedeq1}
\end{equation}

Now, considering the metric (\ref{metric}) with Eq. (\ref{Filedeq1}), one
can write the field equations in the following forms
\begin{eqnarray}
\frac{g(E)^{2}}{d-2}r\psi ^{\prime \prime }(r)+g(E)^{2}\psi ^{\prime }(r)+%
\frac{2}{d\left( d-2\right) }rR_{0} &=&\frac{2^{\frac{d}{4}}\left[ \mathcal{B%
}g(E)f(E)\right] ^{\frac{d}{2}}}{4\left( 1+f_{R}\right) r^{d-1}},  \label{e1}
\\
&&  \nonumber \\
rg(E)^{2}\psi ^{\prime }(r)+\left( d-3\right) g(E)^{2}\left[ \psi (r)-1%
\right] +\frac{rR_{0}}{d} &=&-\frac{2^{\frac{d-4}{4}}\left[ \mathcal{B}%
g(E)f(E)\right] ^{\frac{d}{2}}}{\left( 1+f_{R}\right) r^{d-2}},  \label{e2}
\end{eqnarray}%
which are corresponding to $tt\ $(or $rr$) and $\varphi \varphi $ (or $%
\theta \theta $)\ components, respectively. After some calculations, we can
obtain the metric function in the following form
\begin{equation}
\psi (r)=1-\frac{m}{r^{d-3}}+\frac{2^{\frac{d-4}{4}}\left( \mathcal{B}%
f\left( E\right) g(E)\right) ^{\frac{d}{2}}}{g(E)^{2}\left( 1+f_{R}\right)
r^{d-2}}-\frac{R_{0}r^{2}}{d\left( d-1\right) g\left( E\right) ^{2}}.
\label{g(r)F(R)}
\end{equation}

In order to have well-behaved solutions, hereafter, we restrict ourselves to
$f^{\prime }(R_{0})\neq -1$. We are going to study the general structure of
the solutions. For this purpose, we must investigate the behavior of the
Kretschmann scalar. The Kretschmann scalar goes to infinity ($\infty $) at
the origin ($\lim_{r\longrightarrow 0}R_{\alpha \beta \gamma \delta
}R^{\alpha \beta \gamma \delta }\longrightarrow \infty $). Also the
Kretschmann scalar is finite at infinity ($\lim_{r\longrightarrow \infty
}R_{\alpha \beta \gamma \delta }R^{\alpha \beta \gamma \delta
}\longrightarrow \frac{2}{d\left( d-1\right) }R_{0}^{2}$). So, it shows that
the Kretschmann scalar diverges at $r=0$, and is finite for $r\neq 0$.
Therefore, there is a curvature singularity located at $r=0$. When we define
$R_{0}=\frac{2d}{d-2}\Lambda $, the spacetime will be asymptotically adS. In
order to have physical solutions of this gravity, we should restrict
ourselves to $f_{R}\neq -1$. Also, for $f_{R}>-1$, when we substitute $R_{0}=%
\frac{2d}{d-2}\Lambda $ and $\mathcal{B}\mathbb{=}q[1+f_{R}]^{\frac{2}{d}}$
in Eq. (\ref{g(r)F(R)}), this solution turns into the black hole solution of
Einstein-CIM-gravity's rainbow (\ref{g(r)CIM}) with the same behavior. In
other words, the solutions obtained for $F(R)$ gravity's rainbow in the
presence of CIM source (Eq. (\ref{g(r)F(R)})) are similar to the
Einstein-CIM-rainbow solutions. Therefore, these black hole solutions have
at least one horizon.

\subsection{Thermodynamics}

Here, considering $F(R)$ gravity's rainbow in the presence of CIM source, we
want to obtain the Hawking temperature for the black hole solutions. Using
the definition of surface gravity and Eq. (\ref{g(r)F(R)}), one can
calculate the Hawking temperature as
\begin{equation}
T_{+}=\frac{(1+f_{R})\left[ d(d-3)g(E)^{2}-R_{0}r_{+}^{2}\right]
r_{+}^{d-2}-d\left\{ 2^{d-4}[\mathcal{B}f(E)g(E)]^{2d}\right\} ^{1/4}}{4d\pi
f(E)g(E)r_{+}^{d-1}(1+f_{R})}.  \label{TemF(R)CPMI}
\end{equation}

For this theory, the electric charge and the electric potential of the
conformally invariant $F(R)$ rainbow black hole are, respectively,
\begin{equation}
Q=\frac{d2^{d/4-1}}{16\pi }\left( \frac{\mathcal{B}f(E)}{g(E)}\right)
^{d/2-1},  \label{elcChargeF(R)CPMI}
\end{equation}
\begin{equation}
U=\frac{\mathcal{B}}{r_{+}}.  \label{elcpoF(R)CPMI}
\end{equation}

In the context of modified gravity theories the area law may be generalized
and one can use the Wald entropy associated with the Noether charge \cite%
{Wald}. In order to obtain the entropy of black holes in $F(R)=R+f(R)$
theory, one can use a modification of the area law \cite{Cognola2005}
\begin{equation}
S=\frac{1}{4}\left( \frac{r_{+}}{g(E)}\right) ^{d-2}(1+f_{R}),  \label{SFR}
\end{equation}%
which reveals that the area law does not hold for the obtained black hole
solutions in $F(R)$ gravity's rainbow.

In addition, using the Noether approach, we find that the finite mass can be
obtained as
\begin{equation}
M=\frac{(d-2)\left( 1+f_{R}\right) m}{16\pi f(E)g(E)^{d-3}}.
\label{massF(R)CPMI}
\end{equation}

After calculating the conserved and thermodynamic quantities, we can check
the validity of the first law of black hole thermodynamics. Although $F(R)$
gravity and rainbow functions may modify various quantities, it is
straightforward to show that the modified conserved and thermodynamic
quantities satisfy the first law of thermodynamics as $dM=TdS+UdQ$. The
modified conserved and thermodynamic quantities suggest a deep connection
between the horizon thermodynamics and geometrical properties in modified
gravity.

\section{Thermodynamic Stability of CIM, PMI and F(R) models \label%
{Stability}}

In this section, we employ the canonical ensemble approach toward thermal
stability and phase transition of the solutions. The canonical ensemble is
based on studying the behavior of heat capacity. In this approach, roots of
the heat capacity which are exactly same as the roots of temperature are
denoted as bound points which separate non-physical solutions (having
negative temperature) from physical ones (positive temperature). On the
other hand, the divergencies of the heat capacity are denoted as second
order phase transition points. In other words, in divergence point of the
heat capacity, system goes under a second order phase transition.

The system is in thermal stable state if the signature of heat capacity is
positive. In case of the negative heat capacity, system may acquire
stability by going under phase transition or it may always be unstable which
is known as non-physical case. The heat capacity is obtained as%
\begin{equation}
C_{Q}=\frac{T}{\left( \frac{\partial ^{2}M}{\partial S^{2}}\right) _{Q}}=%
\frac{T}{{\left( \frac{\partial T}{\partial S}\right) _{Q}}}.  \label{CQ}
\end{equation}

Now, considering Eqs. (\ref{TemPMI2}), (\ref{entropyPMI}), (\ref{TemF(R)CPMI}%
) and (\ref{SFR}), one can find following heat capacities%
\begin{equation}
C_{Q}=\left\{
\begin{array}{cc}
-\frac{\left( d-2\right) {r}_{+}^{d-2}\,}{4g^{d-2}\left( E\right) }\left( {%
\frac{\left[ \left( {d}-3\right) g^{2}\left( E\right) -\frac{2\,\Lambda \,{r}%
_{+}^{2}}{d-2}\right] {r}_{+}^{d-3}-\left[ \sqrt{{2}}qf\left( E\right)
g\left( E\right) \right] ^{d-1}}{\left[ \left( {d}-3\right) g^{2}\left(
E\right) +\frac{2\,\Lambda \,{r}_{+}^{2}}{d-2}\right] {r}_{+}^{d-3}-\left(
d-2\right) \left[ \sqrt{{2}}qf\left( E\right) g\left( E\right) \right] ^{d-1}%
}}\right) , & CIM \\
&  \\
-\frac{\left( d-2\right) {r}_{+}^{d-2}\,}{4g^{d-2}\left( E\right) }\left( {%
\frac{\left( {d}-2\right) \left( {d}-3\right) g^{2}\left( E\right) -\,\Theta
\left( 2\,s-\,1\right) {r}_{+}^{2}-2\Lambda {r}_{+}^{2}}{\left( {d}-2\right)
\left( {d}-3\right) g^{2}\left( E\right) -\Theta \left( 2s\,\left[ d-3\,\,%
\right] +1\right) {r}_{+}^{2}+2\Lambda {r}_{+}^{2}}}\right) {,} & PMI \\
&  \\
-\frac{\left( d-2\right) \left( 1+f_{R}\right) {r}_{+}^{d-2}\,}{%
4g^{d-2}\left( E\right) }\left( \frac{\left( 1+f_{R}\right) \left[ \left( {d}%
-3\right) g^{2}\left( E\right) -\frac{2\,\Lambda \,{r}_{+}^{2}}{d-2}\right] {%
r}_{+}^{d-2}-\sqrt[4]{{2}^{d-4}\left( qf\left( E\right) g\left( E\right)
\right) ^{2\,d}}}{\left( 1+f_{R}\right) \left[ \left( {d}-3\right)
g^{2}\left( E\right) +\frac{2\,\Lambda \,{r}_{+}^{2}}{d-2}\right] {r}%
_{+}^{d-2}-\frac{\left( d-1\right) }{\left( d-2\right) }\sqrt[4]{{2}%
^{d-4}\left( qf\left( E\right) g\left( E\right) \right) ^{2\,d}}}\right) , &
F(R)%
\end{array}%
\right. ,
\end{equation}%
in which%
\begin{equation}
\Theta =\left( \frac{2\,\left[ {q}f\left( E\right) g\left( E\right) \left(
2\,s-d+1\right) \right] ^{2}{r}_{+}^{{\frac{4-2\,d}{2\,s-1}}}}{\left(
2\,s-1\right) ^{2}}\right) ^{s}.
\end{equation}

\begin{figure}[tbp]
$%
\begin{array}{ccc}
\epsfxsize=5.5cm \epsffile{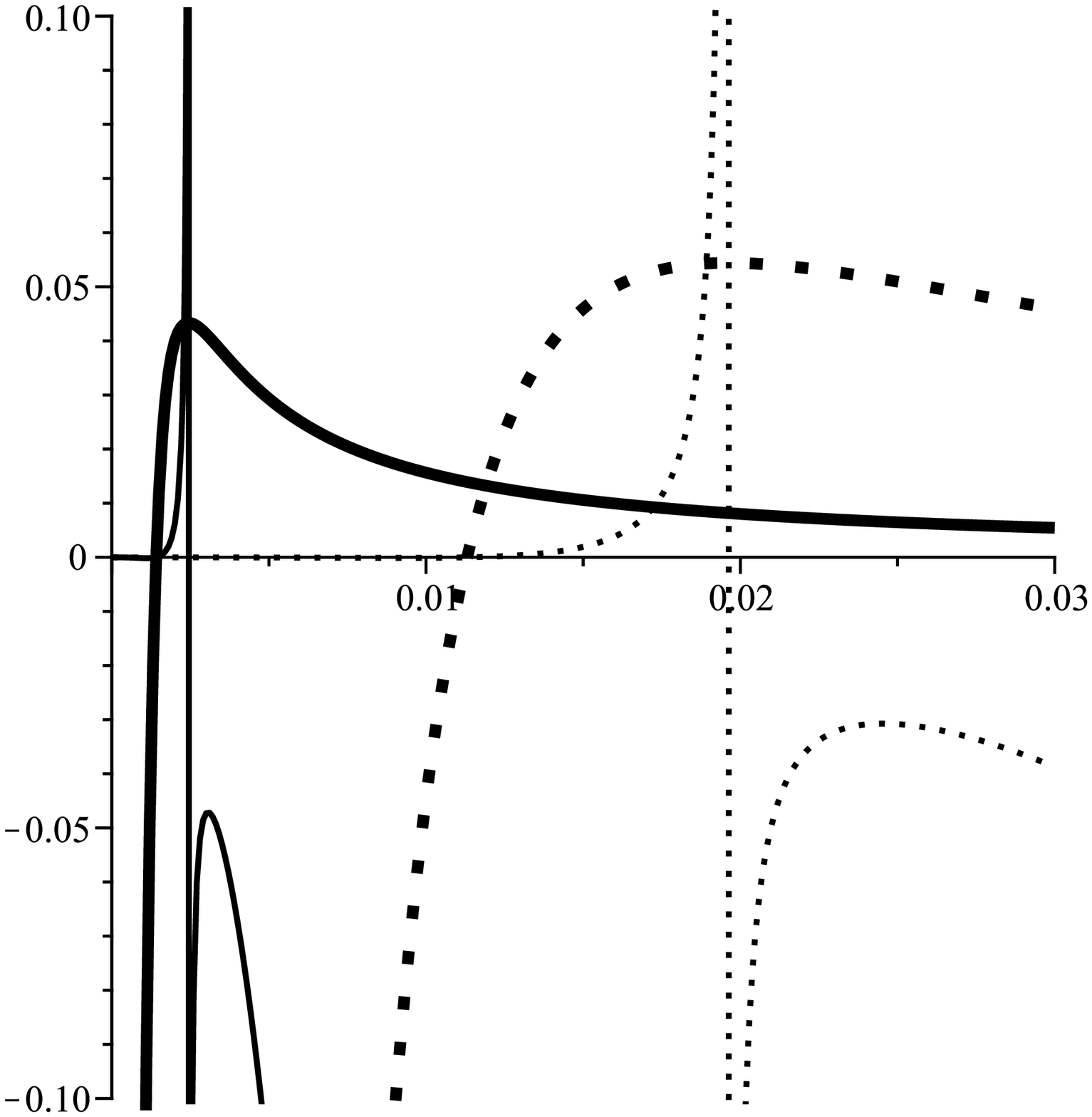} & \epsfxsize=5.5cm %
\epsffile{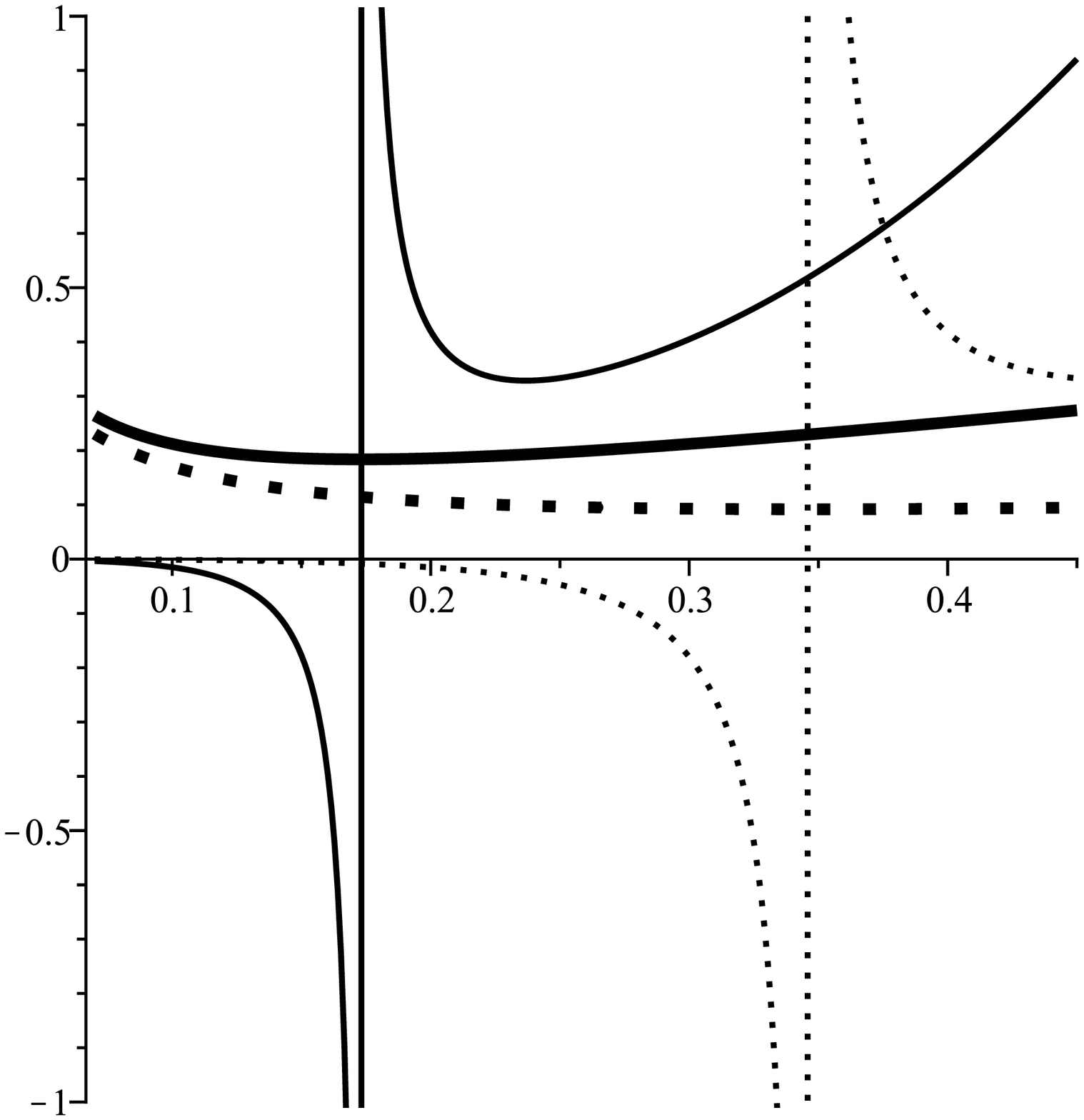} & \epsfxsize=5.5cm %
\epsffile{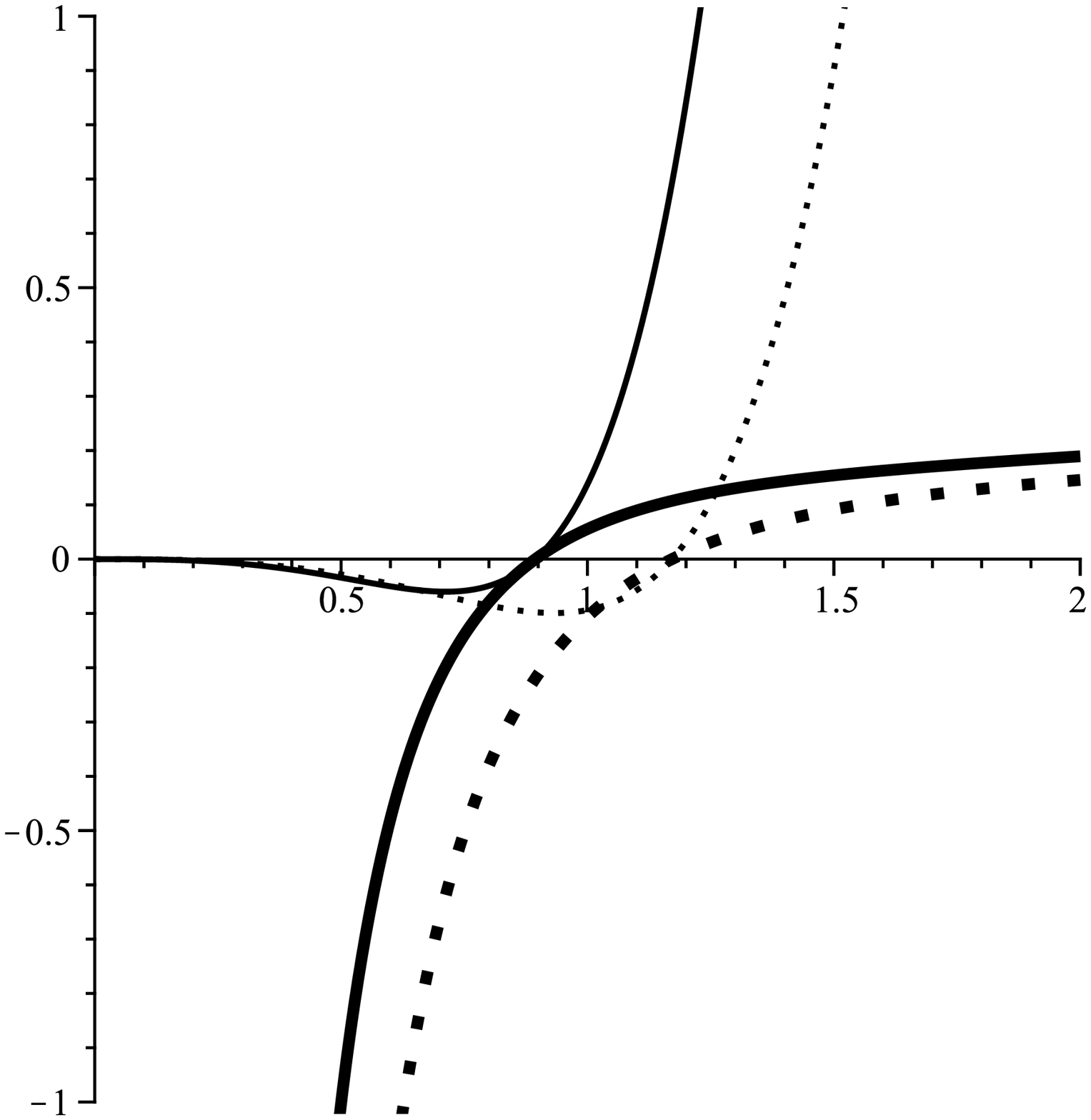}%
\end{array}
$%
\caption{For different scales (CIM case): $C_{Q}$ and $T$ (bold lines)
versus $r_{+} $ for $q=1$, $\Lambda =-1$ and $d=5$;\newline
left and middle panels: $g(E)=f(E)=0.1$ (continues line) and $g(E)=f(E)=0.2$
(dotted line). \newline
right panel: $g(E)=f(E)=0.9$ (continues line) and $g(E)=f(E)=1$ (dotted
line).}
\label{Fig5}
\end{figure}

\begin{figure}[tbp]
$%
\begin{array}{cc}
\epsfxsize=7cm \epsffile{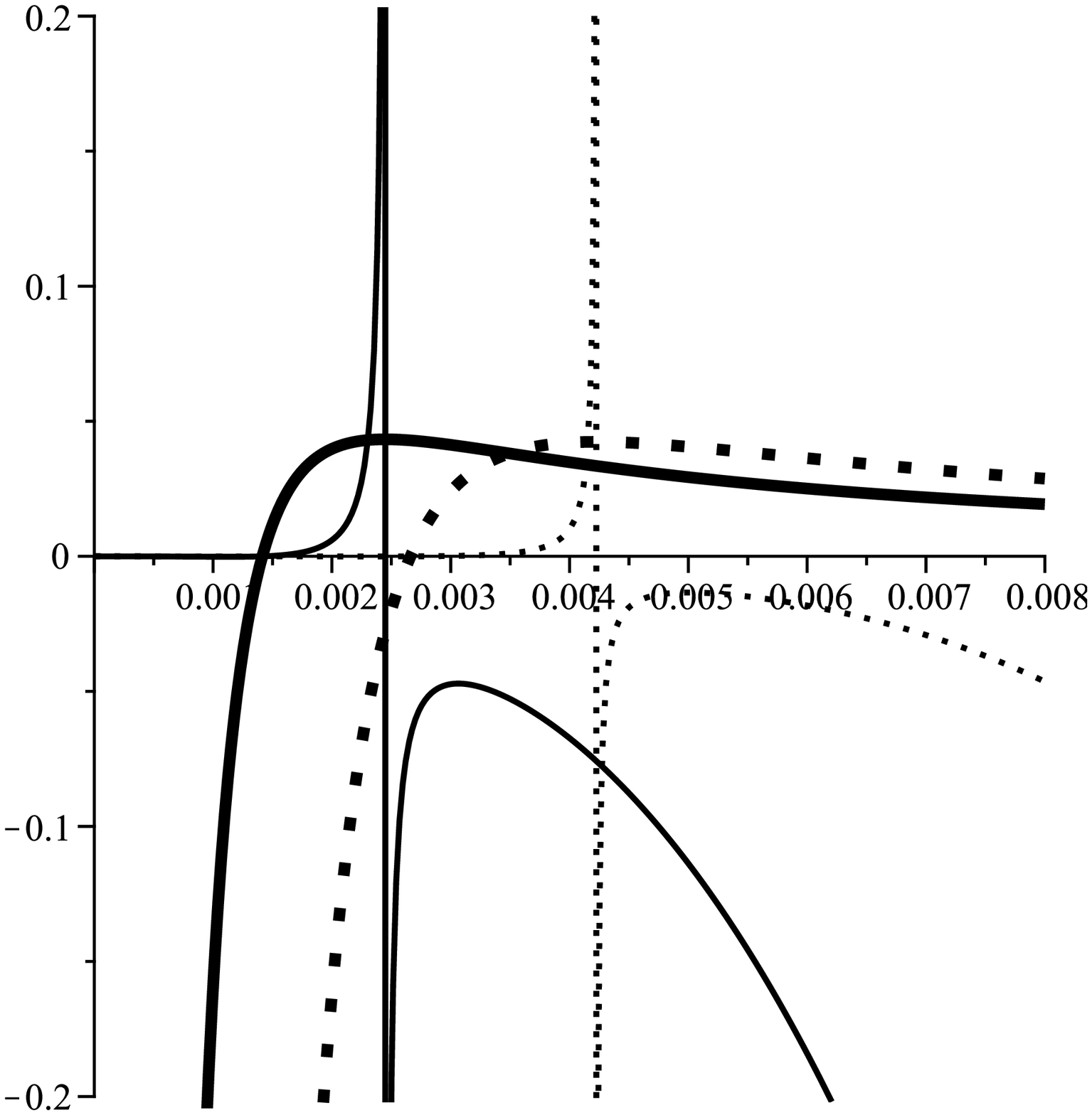} & \epsfxsize=7cm %
\epsffile{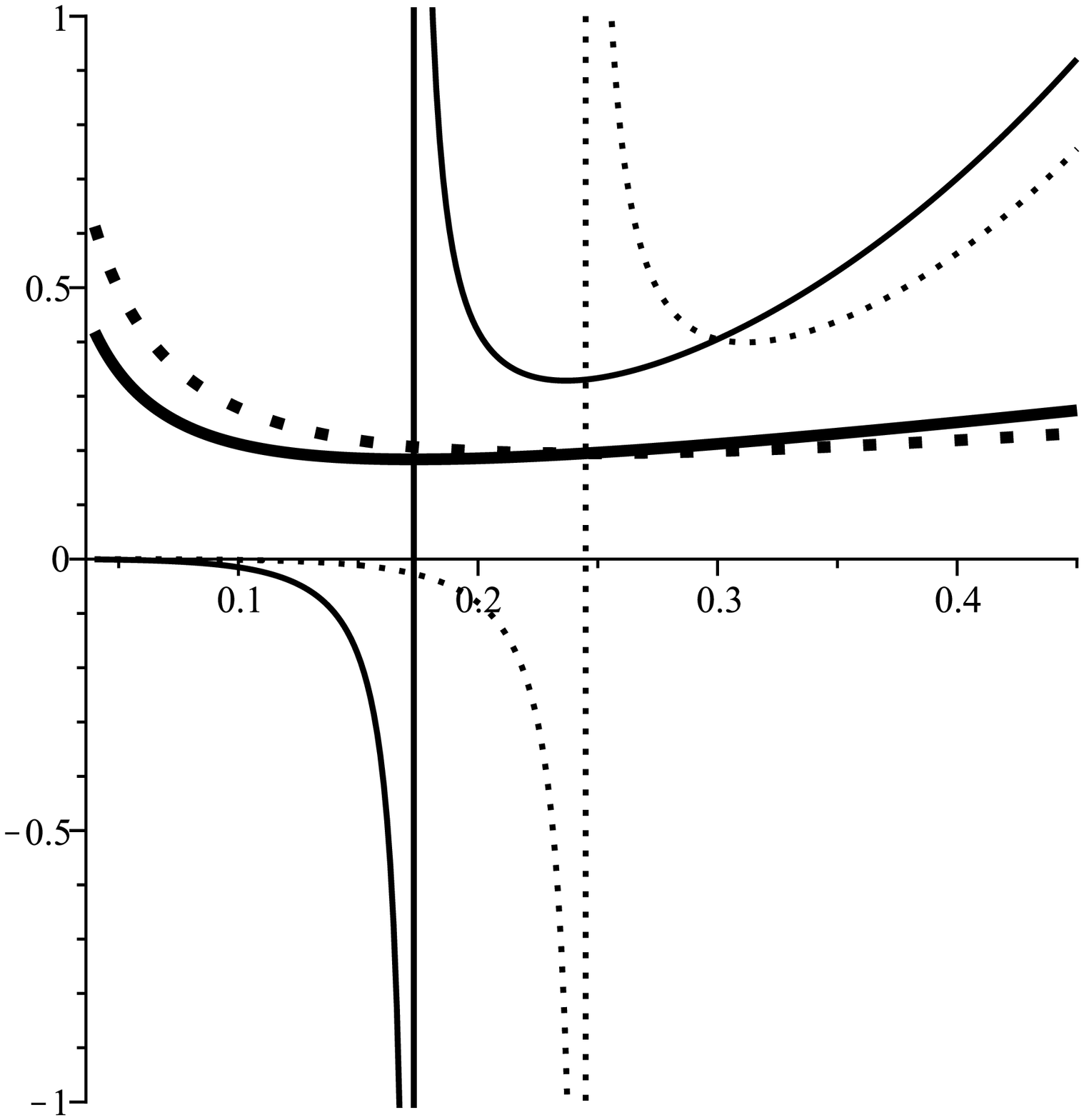}%
\end{array}
$%
\caption{For different scales (CIM case): $C_{Q}$ and $T$ (bold lines)
versus $r_{+} $ for $q=1$, $\Lambda =-1$ and $g(E)=f(E)=0.1$;\newline
$d=5$ (continues line) and $d=6$ (dotted line).}
\label{Fig6}
\end{figure}

\begin{figure}[tbp]
$%
\begin{array}{cc}
\epsfxsize=7cm \epsffile{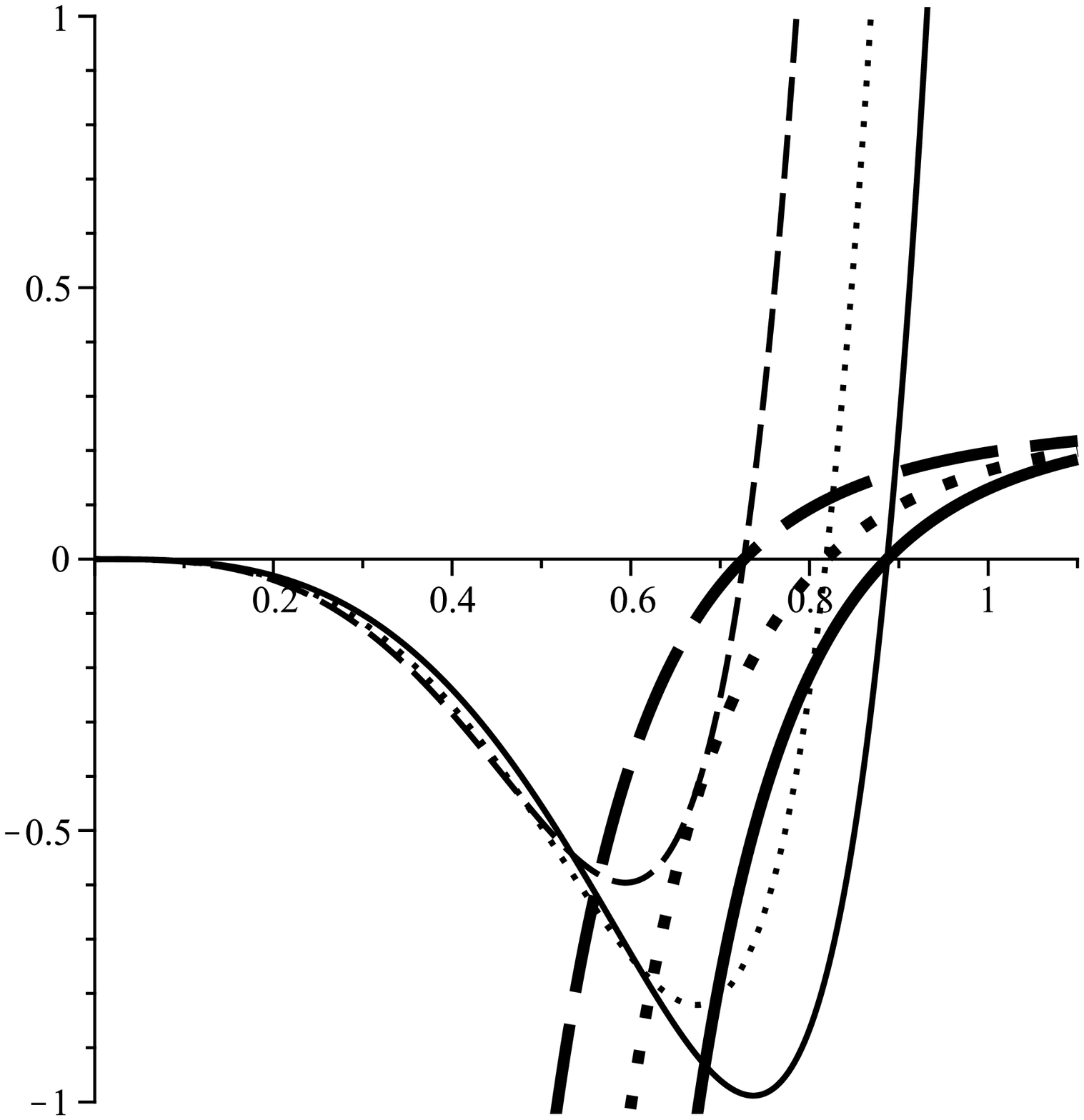} & \epsfxsize=7cm %
\epsffile{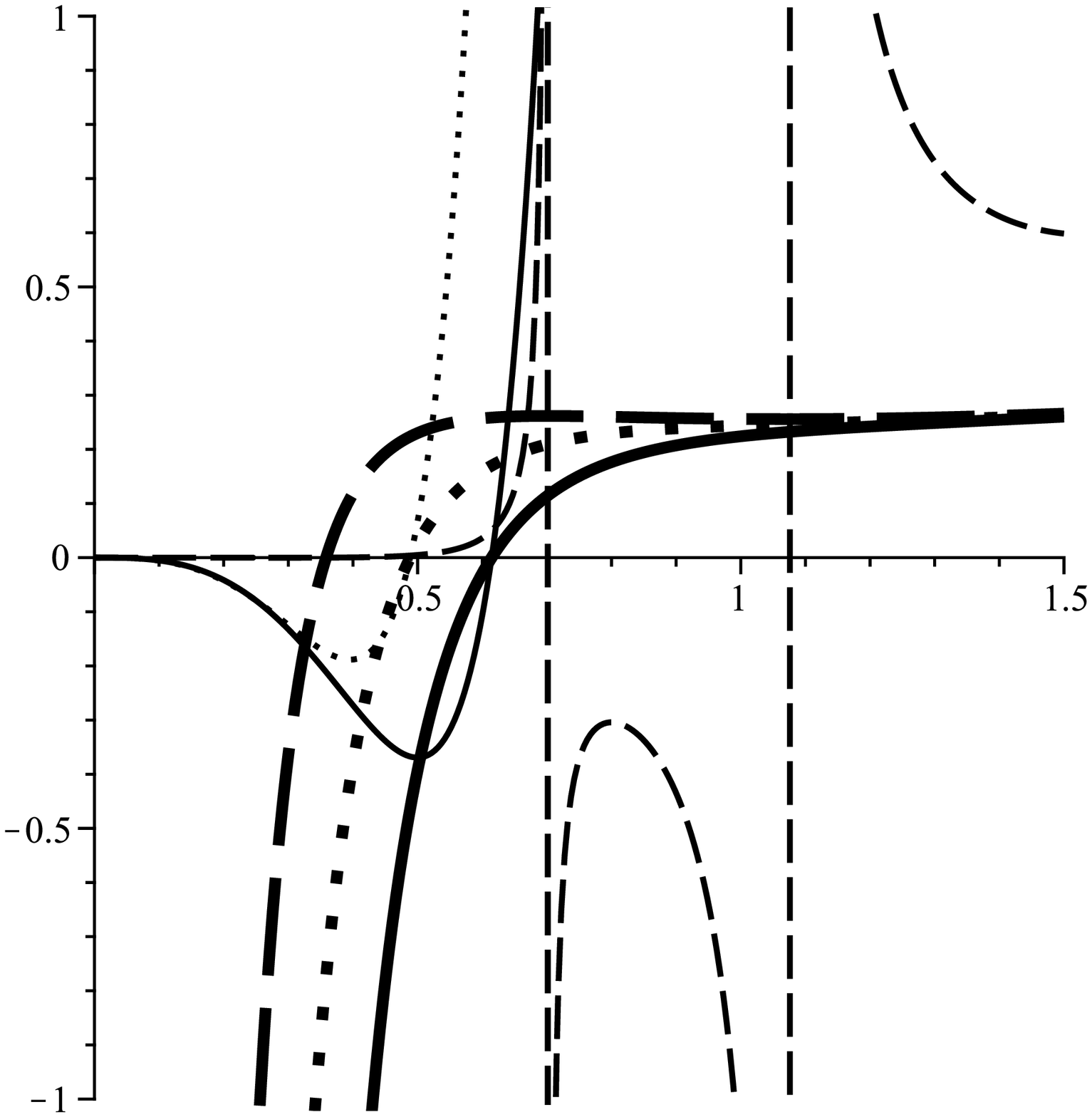}%
\end{array}
$%
\caption{For different scales (PMI case): $C_{Q}$ and $T$ (bold lines)
versus $r_{+} $ for $q=1$, $\Lambda =-1$, $g(E)=f(E)=0.7$ and $d=5$;\newline
left panel: $s=0.9$ (continues line), $s=1$ (dotted line) and $s=1.1$
(dashed line). \newline
right panel: $s=1.2$ (continues line), $s=1.3$ (dotted line) and $s=1.4$
(dashed line).}
\label{Fig7}
\end{figure}

\begin{figure}[tbp]
$%
\begin{array}{cc}
\epsfxsize=7cm \epsffile{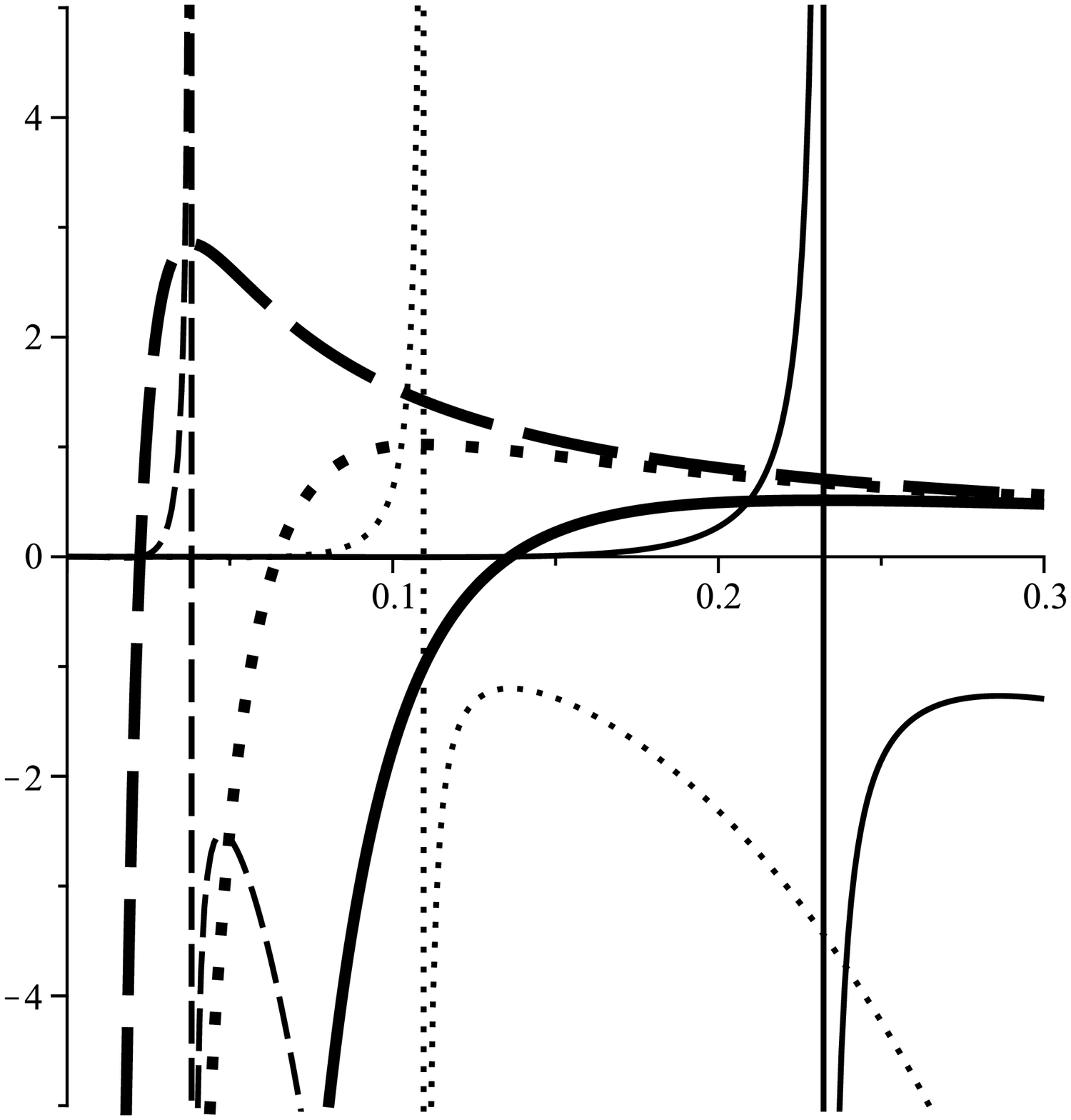} & \epsfxsize=7cm %
\epsffile{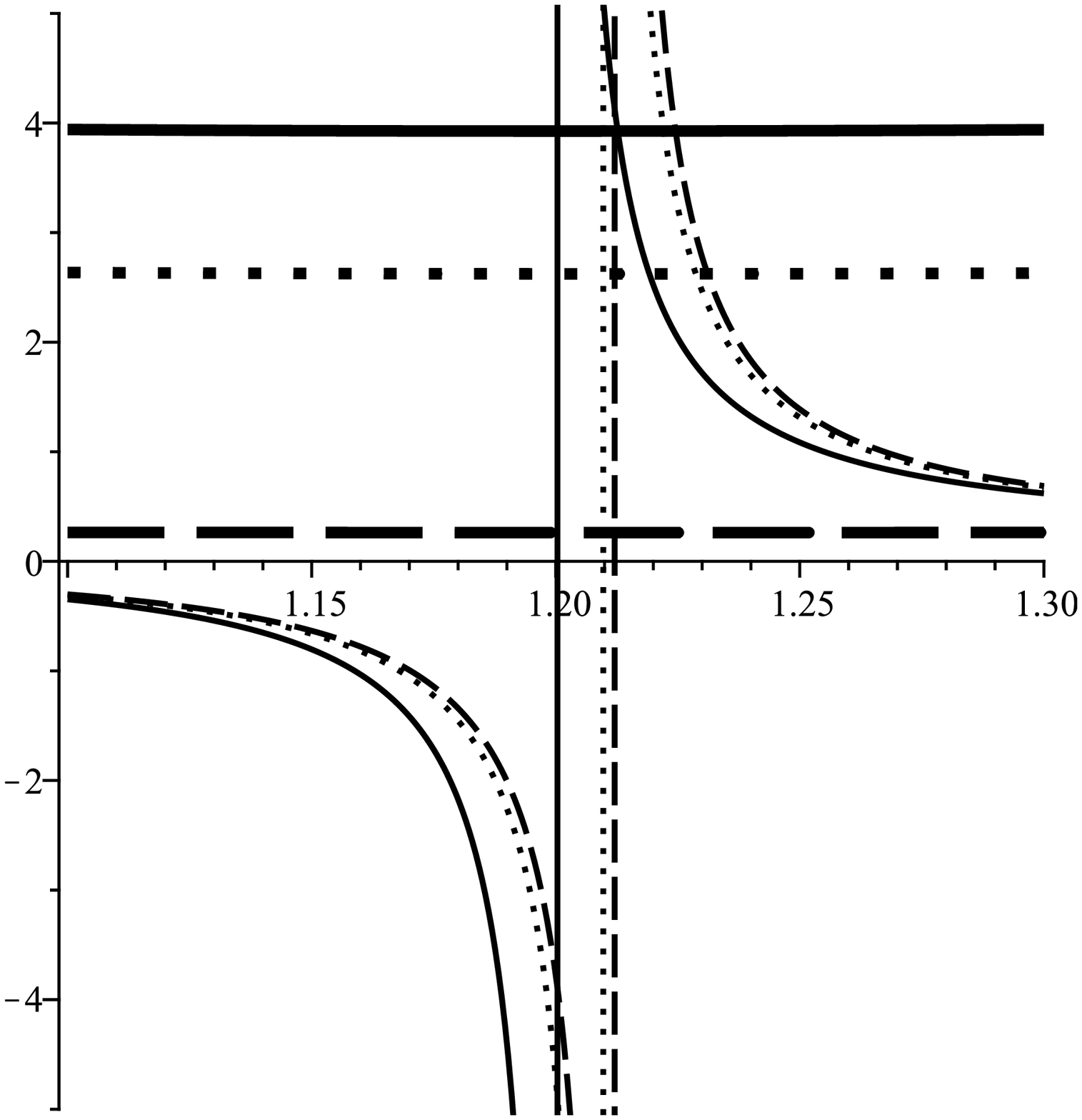}%
\end{array}
$%
\caption{For different scales (PMI case): $C_{Q}$ and $T$ (bold lines)
versus $r_{+} $ for $q=1$, $\Lambda =-1$, $g(E)=f(E)=0.7$ and $d=5$;\newline
$s=1.6$ (continues line), $s=1.7$ (dotted line) and $s=1.8$ (dashed line).}
\label{Fig8}
\end{figure}

\begin{figure}[tbp]
$%
\begin{array}{cc}
\epsfxsize=7cm \epsffile{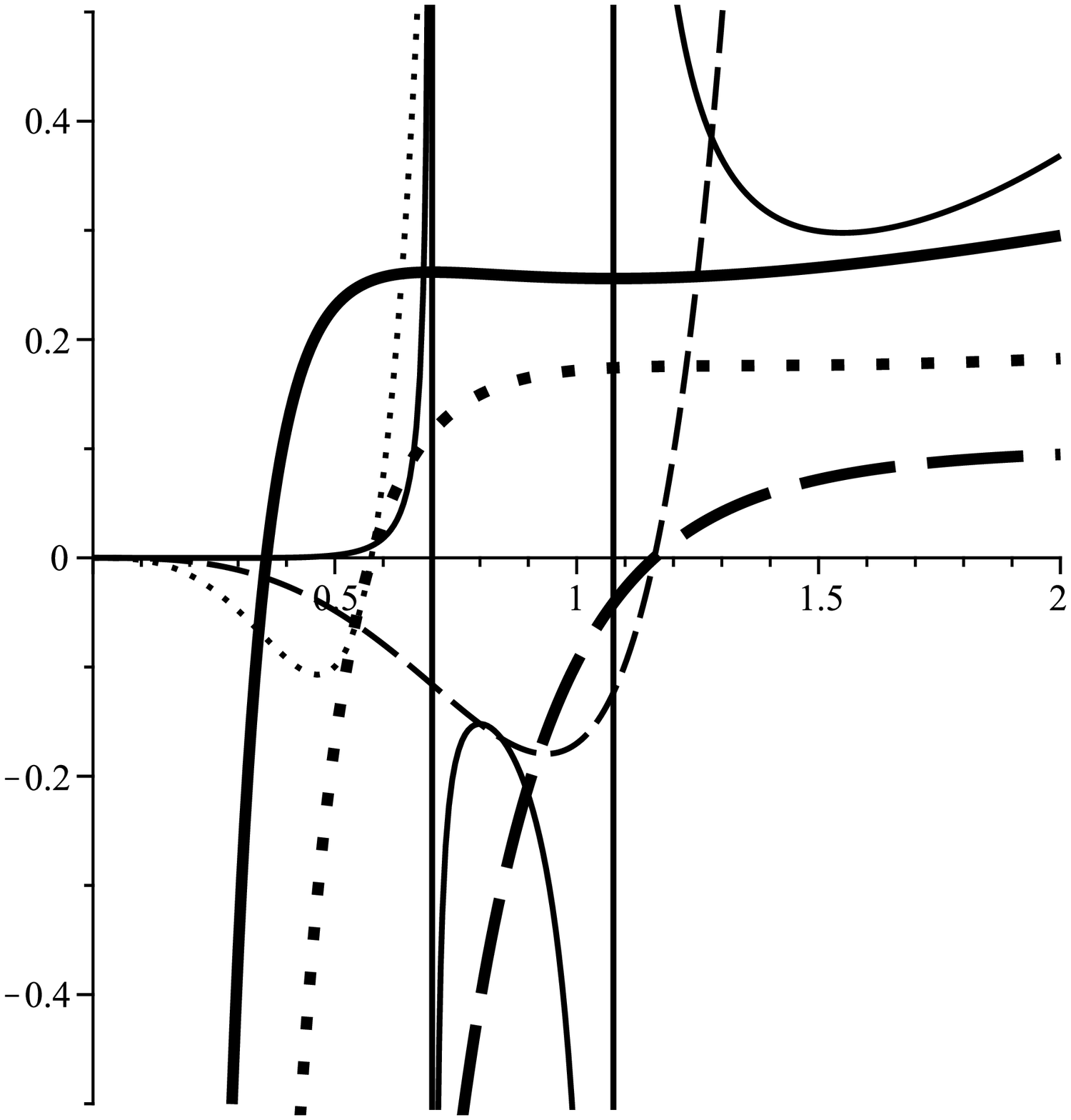} & \epsfxsize=7cm %
\epsffile{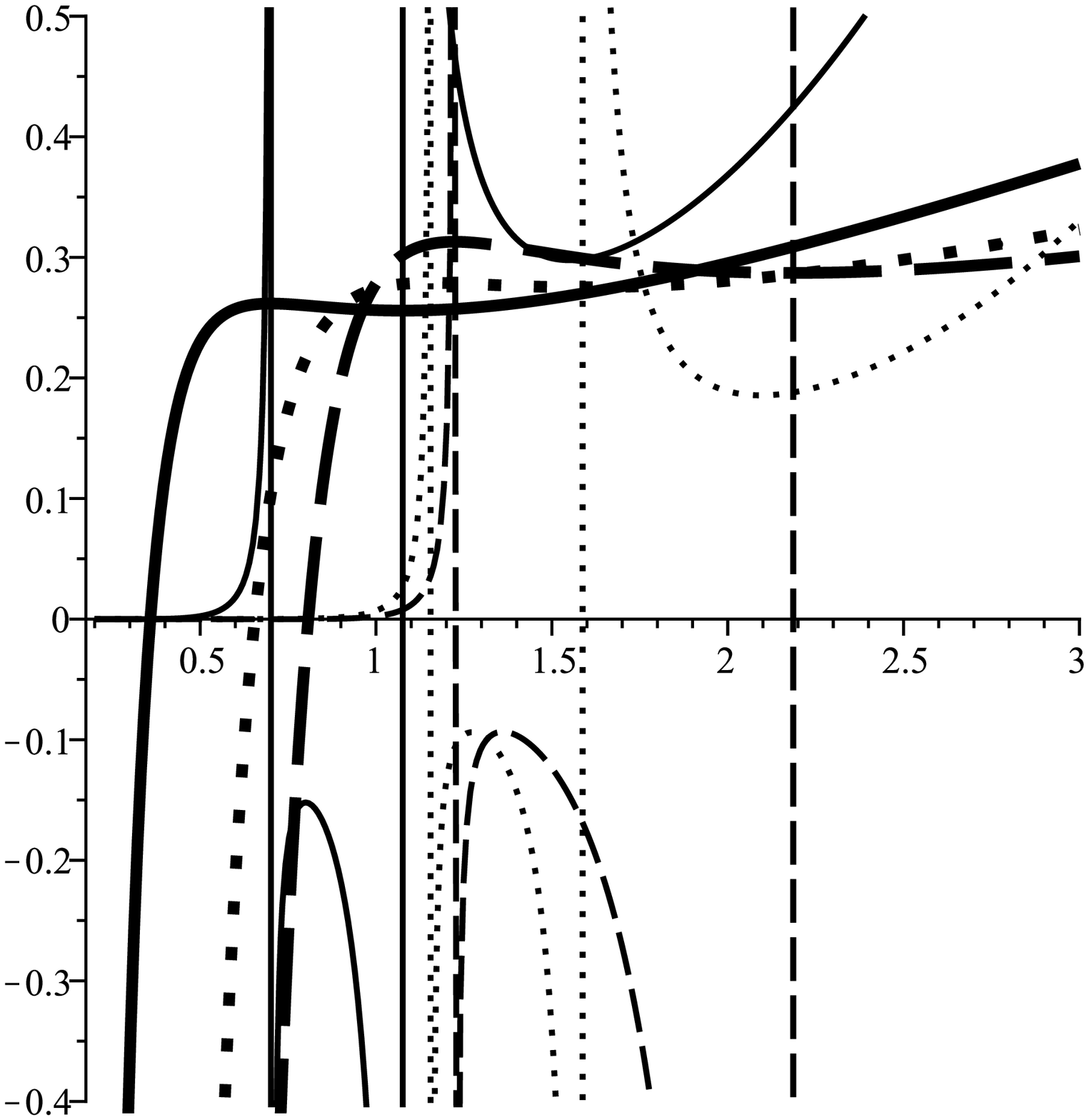}%
\end{array}
$%
\caption{For different scales (PMI case): $C_{Q}$ and $T$ (bold lines)
versus $r_{+} $ for $q=1$, $\Lambda =-1$ and $s=1.4$;\newline
left panel: $d=5$, $g(E)=f(E)=0.7$ (continues line), $g(E)=f(E)=1$ (dotted
line) and $g(E)=f(E)=1.7$ (dashed line). \newline
right panel: $g(E)=f(E)=0.7$, $d=5$ (continues line) and $d=7$ (dotted
line). }
\label{Fig9}
\end{figure}

\begin{figure}[tbp]
$%
\begin{array}{ccc}
\epsfxsize=5.5cm \epsffile{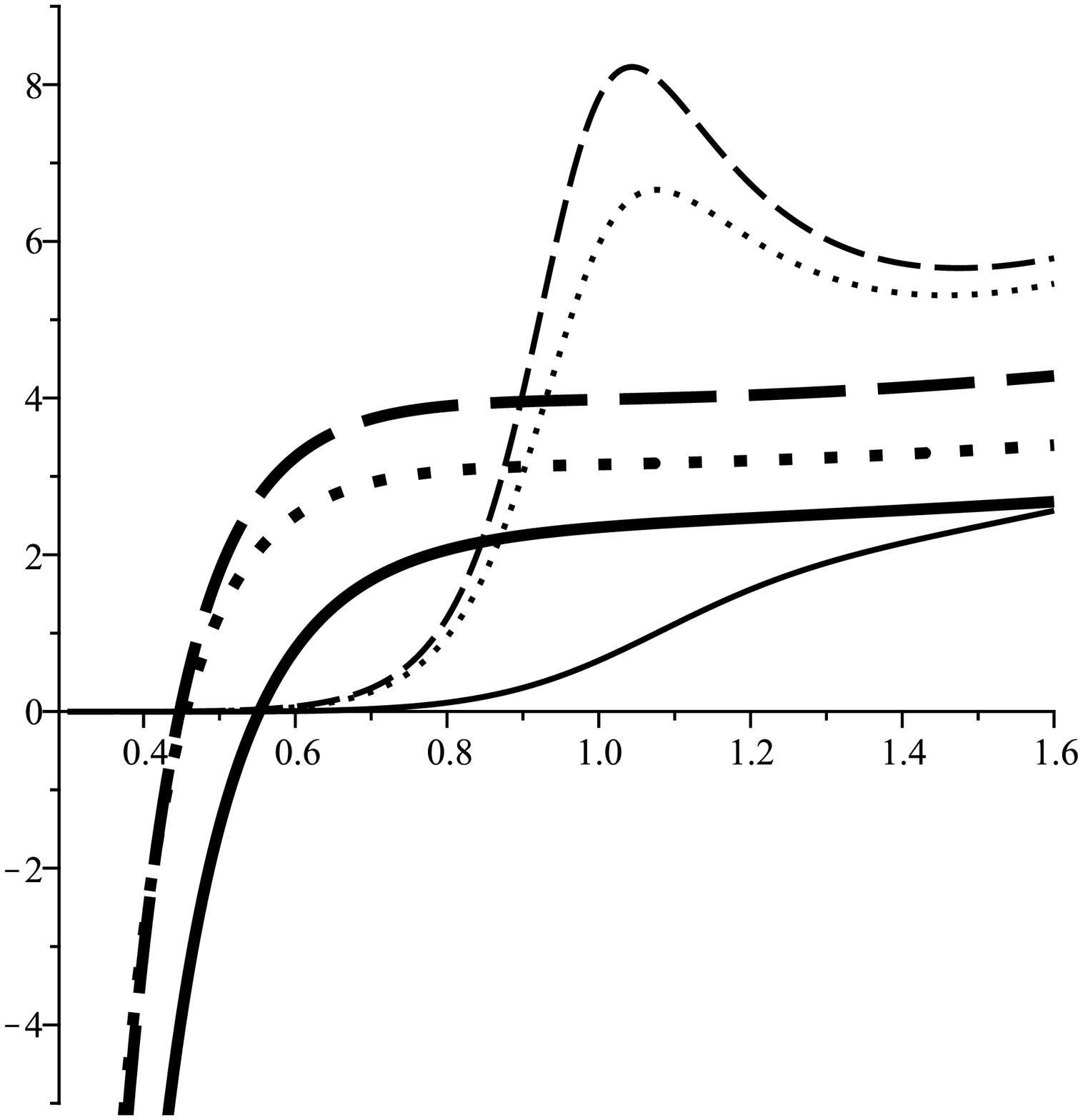} & \epsfxsize=5.5cm %
\epsffile{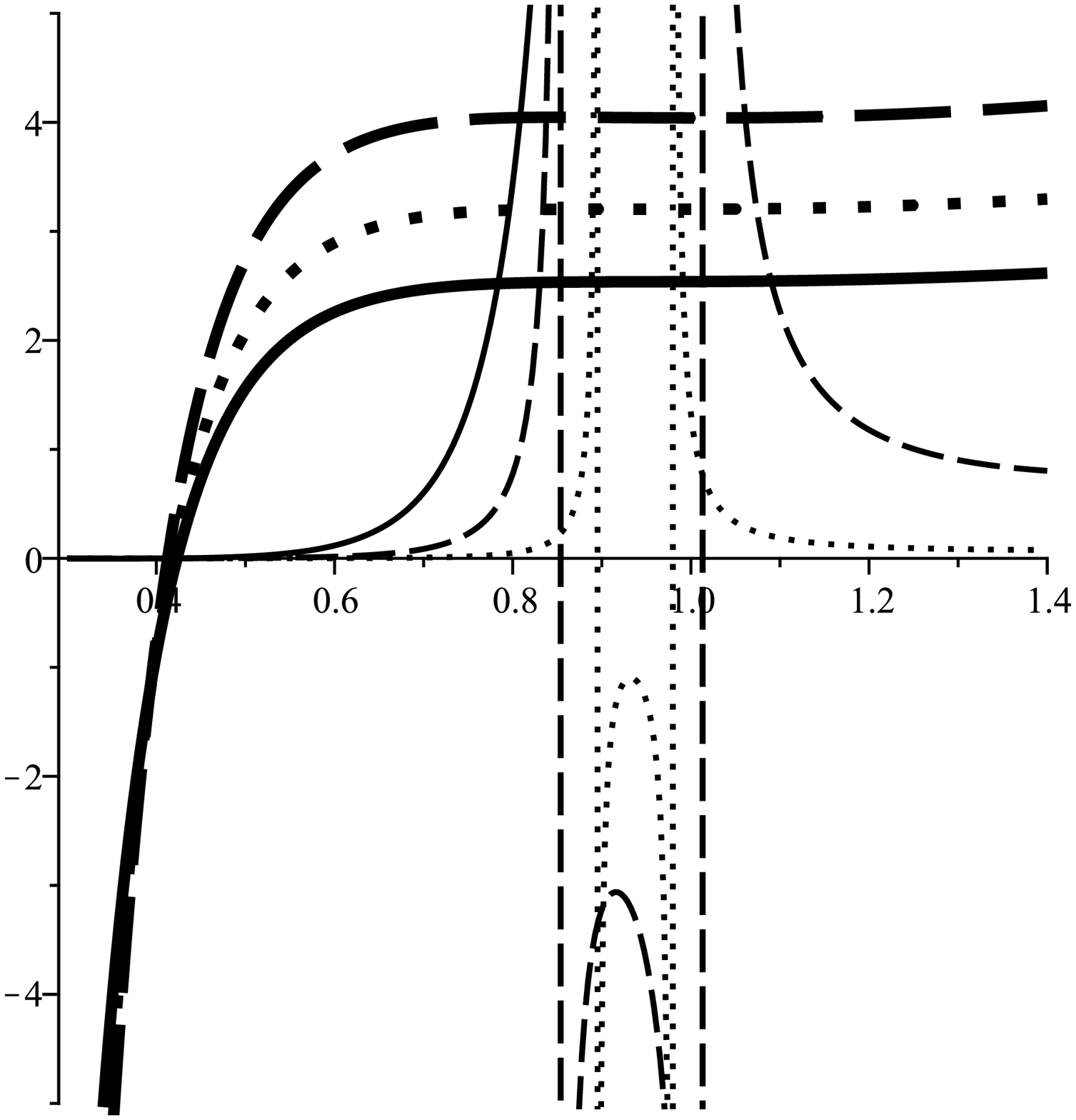} & \epsfxsize=5.5cm \epsffile{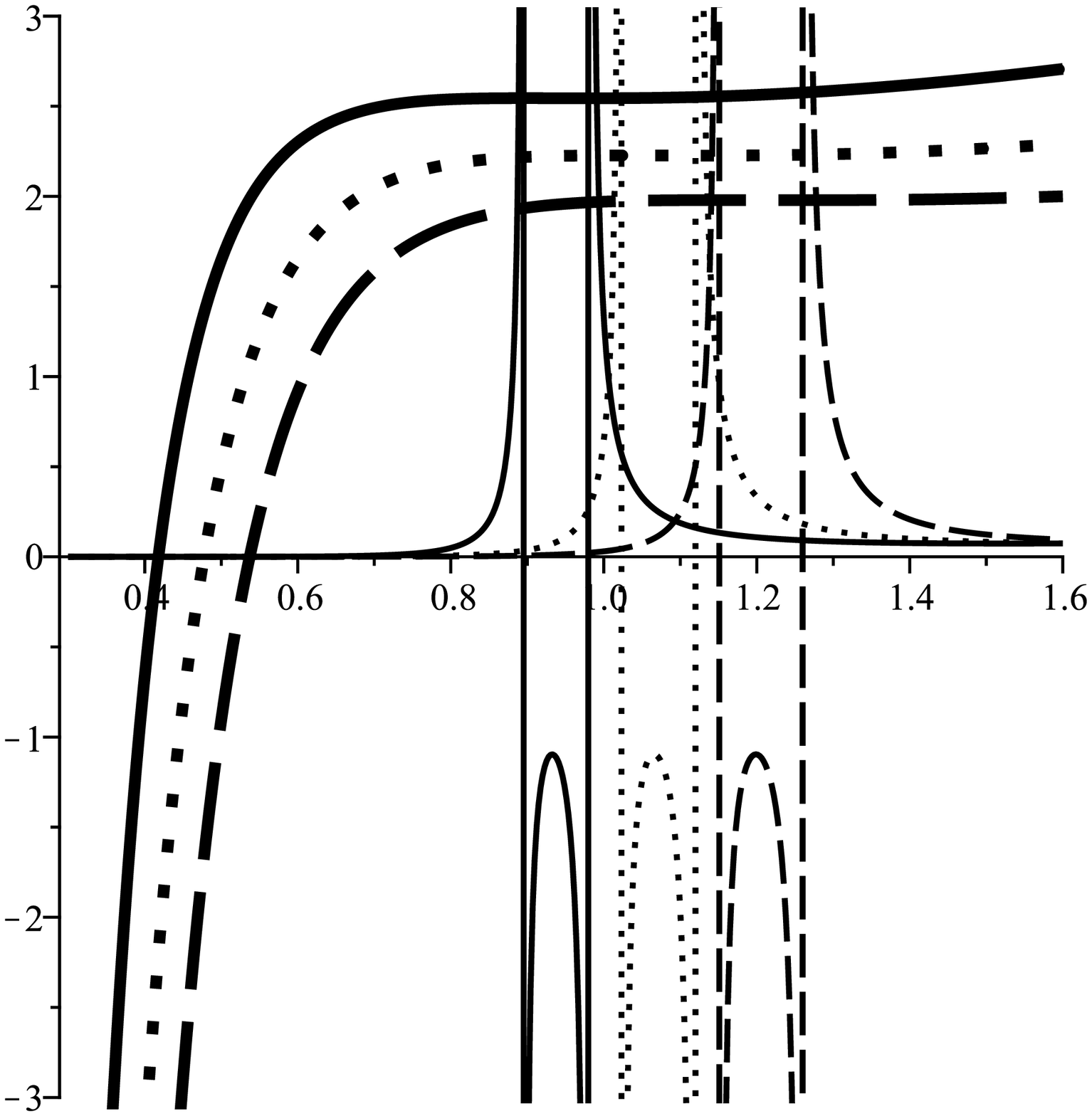}%
\end{array}
$%
\caption{For different scales ($F(R)$ case): $C_{Q}$ and $T$ (bold lines)
versus $r_{+} $ for $q=1$, $\Lambda =-1$ and $d=5$;\newline
left panel: $g(E)=f(E)=0.7$, $f_{R}=0$ (continues line), $f_{R}=0.9$ (dotted
line) and $f_{R}=1$ (dashed line). \newline
middle panel: $g(E)=f(E)=0.7$, $f_{R}=1.4$ (continues line), $f_{R}=1.5$
(dotted line) and $f_{R}=1.6$ (dashed line). \newline
right panel: $f_{R}=1.5$, $g(E)=f(E)=0.7$ (continues line), $g(E)=f(E)=0.8$
(dotted line) and $g(E)=f(E)=0.9$ (dashed line).}
\label{Fig10}
\end{figure}

\begin{figure}[tbp]
$%
\begin{array}{ccc}
\epsfxsize=5.5cm \epsffile{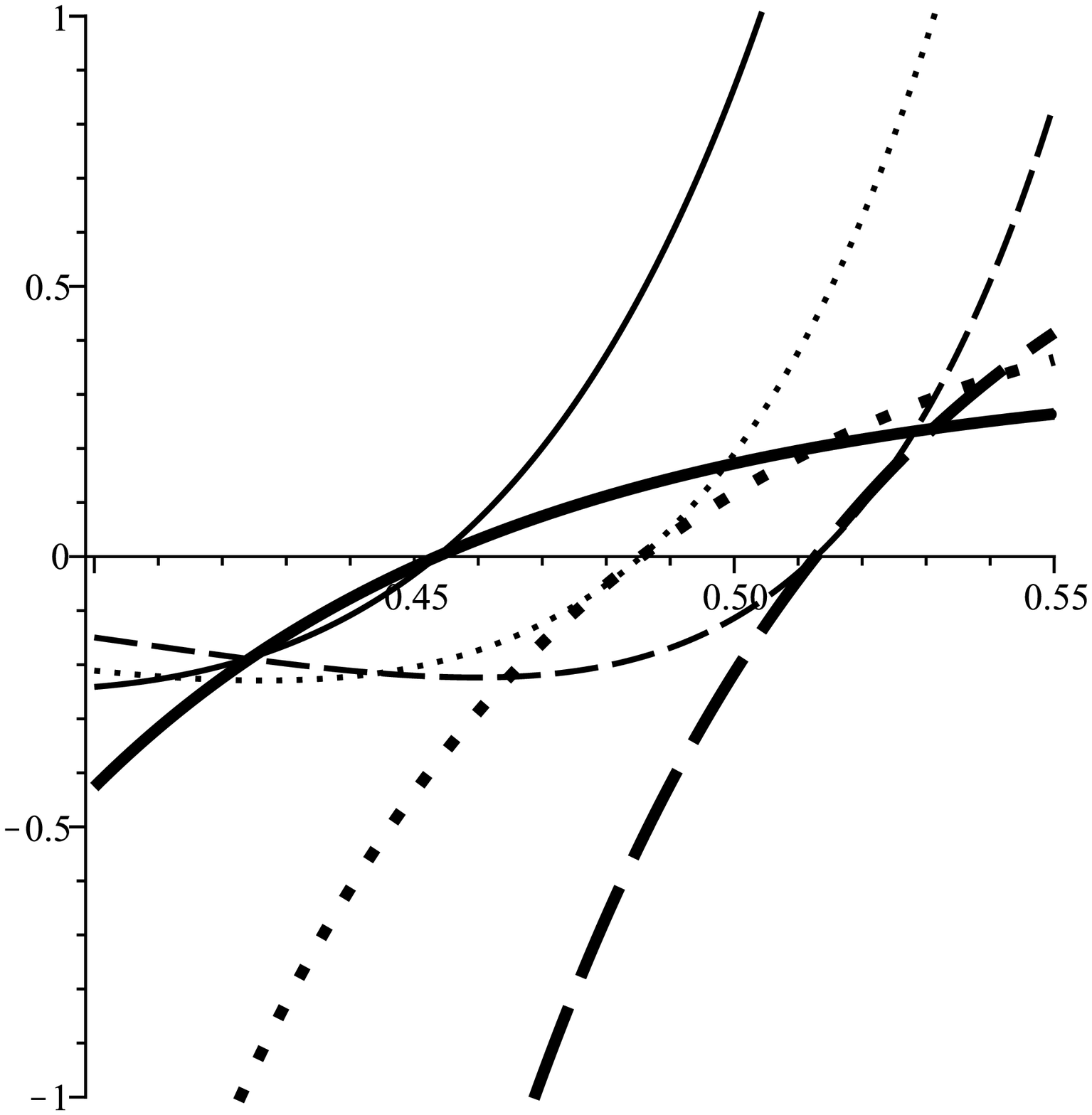} & \epsfxsize=5.5cm %
\epsffile{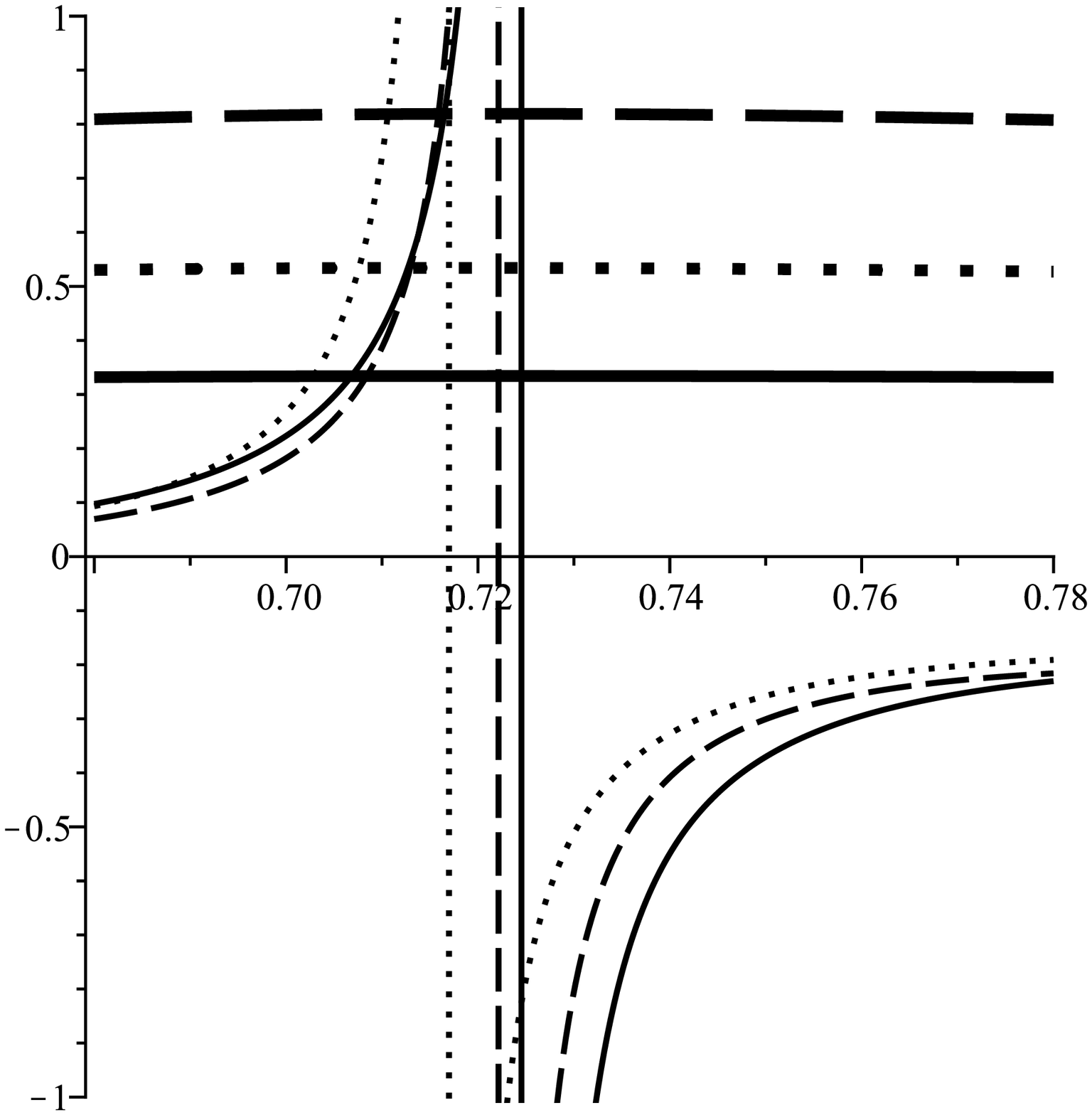} & \epsfxsize=5.5cm \epsffile{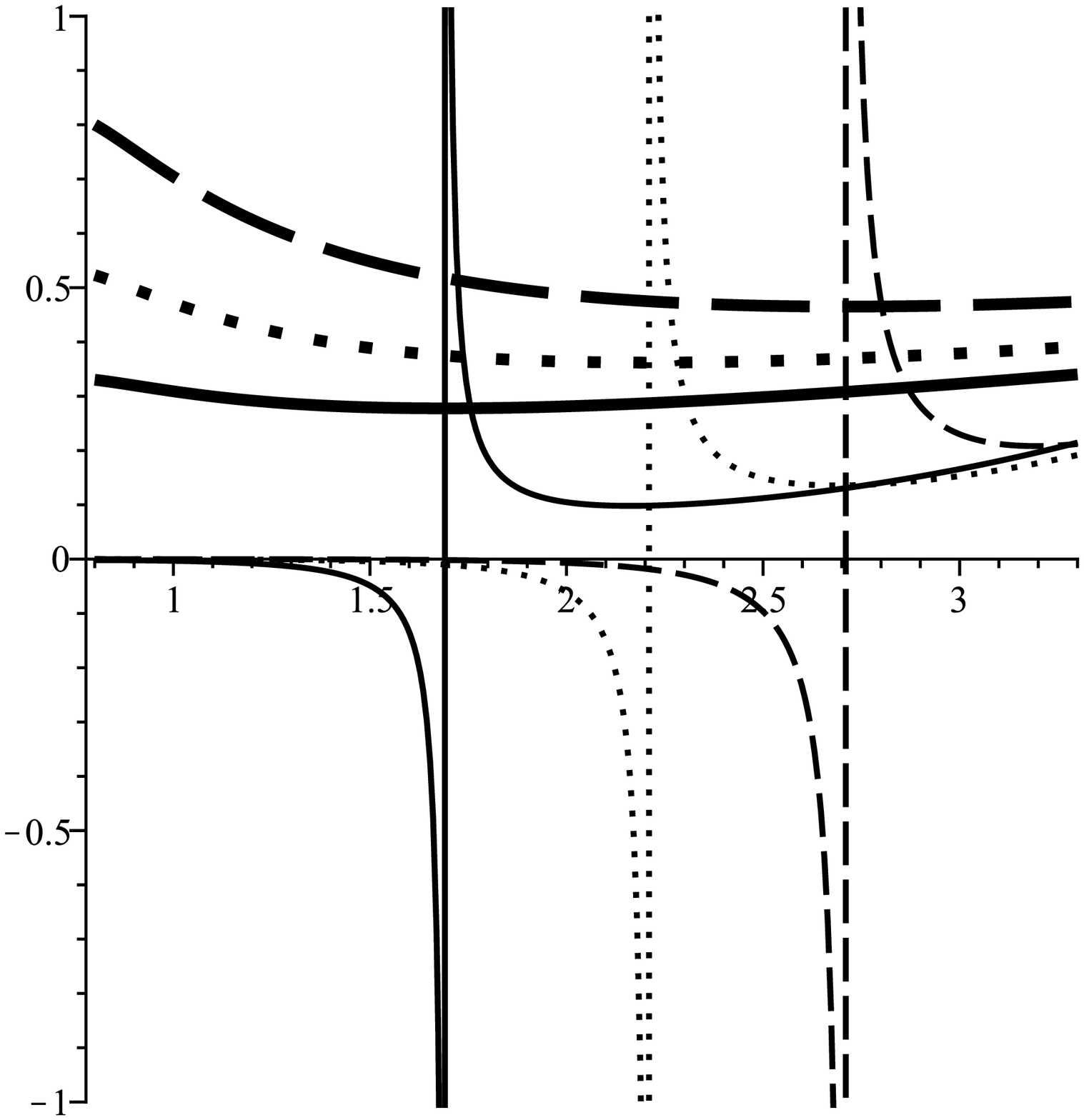}%
\end{array}
$%
\caption{For different scales ($F(R)$ case): $C_{Q}$ and $T$ (bold lines)
versus $r_{+} $ for $q=1$, $\Lambda =-1$, $g(E)=f(E)=0.7$ and $f_{R}=1.5$;%
\newline
left panel: $d=6$ (continues line), $d=7$ (dotted line) and $d=8$ (dashed
line).}
\label{Fig11}
\end{figure}

Due to complexity of the obtained solutions, it is not possible to find the
divergence and bound points analytically, and therefore, we employ numerical
approach. We present the results of our study through different diagrams
(see Figs. \ref{Fig5}-\ref{Fig11}).

Evidently, for CIM and PMI, the thermodynamical stability and phase
transitions are highly sensitive to variations of the energy functions. As
one can see, for these two cases by increasing energy functions, the
stability conditions will be modified completely. For small values of energy
functions, two divergencies and one bound point are observed (left and
middle panels of Fig. \ref{Fig5} and left panel of Fig. \ref{Fig9}). The
stable solutions exist between bound point and smaller divergency and also
after larger divergency. Physical but unstable solutions are between two
phase transition points and otherwise they are non-physical. The
divergencies are marking second order phase transitions. By increasing
energy functions, these behaviors will be modified into two single states of
non-physical unstable and physical stable solutions (right panel of Fig. \ref%
{Fig5} and left panel of Fig. \ref{Fig9}). Interestingly, such effects are
not observed for $F(R)$ gravity. In this case, the variations of energy
functions act as a translation factor. In other words, they shift the places
of bound and divergence points (right panel of Fig. \ref{Fig10}).

As for the PMI solutions, increasing PMI parameter, $s$, leads to
modification in thermal structure of the solutions. For small values of this
parameter, two regions of non-physical unstable and physical stable states
exist (left panel of Fig. \ref{Fig7}). By increasing $s$, a bound point and
two singularities for the heat capacity are formed. The bound point and
smaller divergency are decreasing functions of $s$ while larger divergency
is an increasing function of it (see Figs. \ref{Fig7} and \ref{Fig8}).
Similar behavior is observed for $f_{R}$ in $F(R)$ gravity (left and middle
panels of Fig. \ref{Fig10}). As for the effects of dimensionality, in these
three theories, the bound and divergence points are increasing functions of
this parameter (see Fig. \ref{Fig6}, right panel of Fig. \ref{Fig9} and Fig. %
\ref{Fig11}), the only exception is an interesting behavior for smaller
divergence point for $F(R)$, where an abnormal behavior is observed (see
middle panel of Fig. \ref{Fig11}).

\section{Pure $F(R)$ Gravity's Rainbow}

In this section, we are going to obtain the charged solutions of pure $F(R)$
gravity's rainbow. Therefore, we consider $F(R)=R+f(R)$ gravity without
matter field ($\mathrm{T}_{\mu \nu }^{matt}=0$) for the spherical symmetric
spacetime in $d$-dimensions.

Using the field equation (\ref{EqF(R)1}) and metric (\ref{metric}), one can
obtain the following independent sourceless equations
\begin{eqnarray}
&&F_{R}^{\prime \prime }+\frac{\mathcal{J}_{1}}{2r\psi (r)}F_{R}^{\prime }-%
\frac{\mathcal{J}_{2}}{2r\psi (r)}F_{R}=\frac{F(R)}{2\psi (r)g\left(
E\right) ^{2}},  \label{Fieldtt} \\
&&  \nonumber \\
&&\mathcal{J}_{1}F_{R}^{\prime }-\mathcal{J}_{2}F_{R}=\frac{rF(R)}{g\left(
E\right) ^{2}},  \label{Fieldrr} \\
&&  \nonumber \\
&&F_{R}^{\prime \prime }+\frac{\mathcal{J}_{3}}{r\psi (r)}F_{R}^{\prime }-%
\frac{\left( \mathcal{J}_{3}-d+3\right) }{r^{2}\psi (r)}F_{R}=\frac{F(R)}{%
2\psi (r)g\left( E\right) ^{2}},  \label{Fieldphiphi}
\end{eqnarray}%
where these equations are, respectively, corresponding to $tt$, $rr$ and $%
\varphi \varphi $ components of gravitational field equation. It is notable
that, $F_{R}=\frac{dF(R)}{dR}$, prime and double prime denote

the first and second derivative with respect to $r$, and
\begin{eqnarray*}
\mathcal{J}_{1} &=&r\psi ^{\prime }(r)+2\left( d-2\right) \psi (r), \\
\mathcal{J}_{2} &=&r\psi ^{\prime \prime }(r)+\left( d-2\right) \psi
^{\prime }(r), \\
\mathcal{J}_{3} &=&r\psi ^{\prime }(r)+\left( d-3\right) \psi (r).
\end{eqnarray*}

\subsection{Black hole Solutions}

Here, we study black hole solutions in pure $F(R)$ gravity's rainbow with
constant Ricci scalar ($F_{R}^{\prime \prime }=F_{R}^{\prime }=0$), so it is
easy to show that Eqs. (\ref{Fieldtt})-(\ref{Fieldphiphi}) reduce to the
following forms
\begin{eqnarray}
g\left( E\right) ^{2}\left[ r\psi ^{\prime \prime }(r)+\left( d-2\right)
\psi ^{\prime }(r)\right] F_{R} &=&-rF(R),  \label{eqaa} \\
2g\left( E\right) ^{2}\left[ r\psi ^{\prime }(r)+\left( d-3\right) \left(
\psi (r)-1\right) \right] F_{R} &=&-r^{2}F(R).  \label{eqbb}
\end{eqnarray}

It is notable that, there are different models of $F(R)$ gravity which may
explain some of local (or global) properties of the universe. One of these
models which supposed to explain the positive acceleration of expanding
universe was $F(R)=R-\frac{\mu ^{4}}{R}$ model \cite{Capozziello5}. Another
model of $F(R)$ gravity which is introduced by Kobayashi and Maeda \cite%
{Kobayashi} is $F(R)=R+\kappa R^{n}$ (where $n>1$). This model can resolve
the singularity problem arising in the strong gravity regime. On the other
hand, by adding an exponential correction term to the Einstein Lagrangian
\cite{FR3,NojiriO,Elizalde}, one can prove that this model can satisfy both
Solar system tests and high curvature condition \cite{Zhang}. Also, one of
the interesting models that passes all the theoretical and observational
constraints is known as the Starobinsky model (its functional form is $%
F(R)=R+\lambda R_{0}\left( \left( 1+\frac{R^{2}}{R_{0}^{2}}\right)
^{-n}-1\right) $, \cite{StarobinskyS,StarobinskyJ}). It is notable that,
this model could produced viable cosmology different from the $\Lambda$CDM
one at recent times and also satisfy Solar system and cosmological tests,
simultaneously. Moreover, its is worthwhile to mention that there are
various models of $F(R)$ gravity in which these models have interesting
properties for describing our universe and also Solar system (see \cite%
{ZhangP,Appleby,Hu,AmendolaG,Tsujikawa,Linder,NojiriOS,BambaGL,Girones,He,Huang,OdintsovO,BambaOS,KruglovS,Kusakabe,Geng,Dutta}%
, for an incomplete list of references in this directions).

Besides, another interesting model of $F(R)$ gravity was introduced by Bamba
et al. \cite{BambdaNOS}. They constructed an $F(R)$ gravity theory
corresponding to the Weyl invariant two scalar field theory ($F(\overline{R}%
)=\frac{e^{\eta \left( \varphi (\overline{R})\right) }}{12}\left[ 1-\varphi (%
\overline{R})\right] \overline{R}-e^{2\eta \left( \varphi (\overline{R}%
)\right) }J(\varphi (\overline{R}))$, see Ref. \cite{BambdaNOS}, for more
details). Their model can have the antigravity regions in which the Weyl
curvature invariant does not diverge at the Big Crunch and Big Bang
singularities. Also, Nojiri and Odintsov investigated the anti-evaporation
of Schwarzschild-de Sitter and Reissner-Nordstr\"{o}m black holes by several
interesting $F(R)$ models such as; $F(R)=\frac{R}{2\kappa ^{2}}%
+f_{0}M^{4-2n}R^{n}+f_{2}R^{2}$ \cite{NojiriOI} and $F(R)=\frac{R}{2\kappa
^{2}}\left( 1-\frac{R}{R_{0}}\right) $ \cite{NojiriOII}.

In order to obtain an exact solution, we should choose a proper model of $%
F(R)$. In this regard, we use the following interesting models of $F(R)$
gravity with appropriate properties
\begin{equation}
\begin{array}{cc}
F(R)=R-\lambda e^{-\xi R}+\eta R^{n}, & \text{type}-I \\
&  \\
F(R)=R+\alpha R^{n}-\beta R^{2-n}, & \text{type}-II \\
&  \\
F(R)=R-m^{2}\frac{c_{1}\left( \frac{R}{m^{2}}\right) ^{n}}{c_{2}\left( \frac{%
R}{m^{2}}\right) ^{n}+1}, & \text{type}-III \\
&  \\
F(R)=R-a\left[ e^{-bR}-1\right] +cR^{N}\frac{e^{bR}-1}{e^{bR}+e^{bR_{0}}}, &
\text{type}-IV%
\end{array}%
.
\end{equation}

\textbf{Type}$-I$\textbf{:} This model includes an additional exponential
term ($\lambda e^{-\xi R}$). It was shown that such model enjoys validity of
the Solar system tests and high curvature condition \cite{Zhang} whereas it
suffers instability for its solutions. In order to remove such instability,
it is possible to add a correction term ($\eta R^{n}$ with $n>1$) such as
one introduced by Kobayashi and Maeda \cite{Kobayashi}, in which the
singularity problem in strong gravity regime is removed. It is worthwhile to
mention that for $n=2$, this model can be used to explain inflation
mechanism \cite{NojiriO,Elizalde,StarobinskyS}. The type-$I$ model and some
of its properties have been investigated in \cite{HendiGRG,HendiES}.

\textbf{Type}$-II$\textbf{:} Considering a scalar potential with non-zero
value of residual vacuum energy, another model of $F(R)$ gravity was
proposed by Artymowskia and Lalak \cite{Artymowski}. The reason for such
consideration was to have a source for dark energy. This model is a stable
minimum of the Einstein frame scalar potential of the auxiliary field. The
results of this model are consistent with PLANCK data and also lack of
eternal inflation. It should be pointed out that here, $\alpha$ and $\beta$
are positive constants and $n$ is within the range $\left[ \frac{1}{2}\left(
1+\sqrt{3}\right) ,2\right]$ (see Refs. \cite{FR7,Artymowski}, for more
details).

\textbf{Type}$-III$\textbf{:} The third model was introduced by Hu and
Sawicki \cite{Hu}. This model provides a description regarding accelerated
expansion without cosmological constant. In addition, this model enjoys the
validity of both cosmological and Solar-system tests in the small-field
limit. It should be pointed out that, $n$ in this model is positive valued ($%
n>0$) where for the case $n=1$, it covers the mCDTT with a cosmological
constant at high curvature regimes. Furthermore, for $n=2$, the inverse
curvature squared model is included \cite{Mena} with a cosmological
constant. In this model, $c_{1}$ and $c_{2}$ are dimensionless parameters
and $m^{2}$ is the mass scale (for more details, see Ref. \cite{Hu}).

\textbf{Type}$-IV$\textbf{:} The last model is able to describe both
inflation in the early universe and the recent accelerated expansion \cite%
{FR3}. In addition, this model pass all local tests such as stability of
spherical body solutions, generation of a very heavy positive mass for the
additional scalar degree of freedom and non-violation of Newton's law. In
this model, $N>2$ and $c$ is a positive quantity (see Ref. \cite{FR3}, for
more details).

To find the metric function $\psi(r)$, we use the components of Eqs. (\ref%
{eqaa}) and (\ref{eqbb}) with the models under consideration. Using Eqs. (%
\ref{eqaa}) and (\ref{eqbb}) with the metric (\ref{metric}), one can obtain
a general solution for these pure gravity models in the following form
\begin{equation}
\psi (r)=1-\frac{2\Lambda r^{2}}{\left( d-1\right) \left( d-2\right) g\left(
E\right) ^{2}}-\frac{m}{r^{d-3}}+\frac{2^{\frac{d-4}{4}}\left[ qf\left(
E\right) g(E)\right] ^{\frac{d}{2}}}{g(E)^{2}r^{d-2}}.  \label{g(r)pure}
\end{equation}

In order to satisfy all components of the field equations (Eqs. (\ref{eqaa})
and (\ref{eqbb})), we should set the parameters of $F(R)$ model such that
the following equations are satisfied
\begin{equation}
\begin{array}{c}
\text{type}-I\text{:}\left\{
\begin{array}{c}
R\left[ (n-\frac{d}{2})\eta R^{n-1}-\frac{d-2}{2}\right] e^{\xi R}+\lambda
\left( \frac{d}{2}+\xi R\right) =0, \\
\\
\left( 1+n\eta R^{n-1}\right) e^{\xi R}+\lambda \xi =0,%
\end{array}%
\right. , \\
\\
\text{type}-II\text{:}\left\{
\begin{array}{c}
\left[ \alpha \left( n-\frac{d}{2}\right) R^{n}-\frac{d-2}{2}R\right]
R^{n}+\left( n+\frac{d-4}{2}\right) \beta R^{2}=0, \\
\\
\left( \alpha nR^{n}+R\right) R^{n}+\left( n-2\right) \beta R^{2}=0,%
\end{array}%
\right. , \\
\\
\text{type}-III\text{:}\left\{
\begin{array}{c}
c_{2}\left[ \frac{d}{d-2}m^{2}c_{1}-c_{2}R\right] \left( \frac{R}{m^{2}}%
\right) ^{2n}-\left[ \frac{2m^{2}c_{1}}{d-2}\left( n-\frac{d}{2}\right)
+2c_{2}R\right] \left( \frac{R}{m^{2}}\right) ^{n}-R=0, \\
\\
\left[ c_{2}^{2}\left( \frac{R}{m^{2}}\right) ^{2n}+1\right] R+\left[
2c_{2}R-m^{2}nc_{1}\right] \left( \frac{R}{m^{2}}\right) ^{n}=0,%
\end{array}%
\right. , \\
\\
\text{type}-IV\text{:}\left\{
\begin{array}{c}
\left[ c\left( N-\frac{d}{2}\right) R^{N}-\frac{d-2}{2}R-\frac{d}{2}a\right]
e^{3bR}+a\left( bR+\frac{d}{2}\right) e^{2bR_{0}} \\
+\left\{ c\left[ \left( bR+N-\frac{d}{2}\right) e^{bR_{0}}+bR-N+\frac{d}{2}%
\right] R^{N}\right.  \\
\left. -\left[ \left( d-2\right) R+ad\right] e^{bR_{0}}+a\left( bR+\frac{d}{2%
}\right) \right\} e^{2bR}+ \\
\left. \left[ a\left( 2bR+d\right) -c\left( N-\frac{d}{2}\right)
R^{N}-\left( \frac{ad}{2}-\frac{\left( d-2\right) R}{2}\right) e^{bR_{0}}%
\right] e^{b\left( R+R_{0}\right) }\right. =0, \\
\\
\begin{array}{c}
\left( 1+cNR^{N-1}\right) e^{3bR}+\left[ e^{bR_{0}}\left(
2ab+e^{bR_{0}}\right) -cNR^{N-1}e^{bR_{0}}\right] e^{bR} \\
\left. +\left[ cR^{N-1}\left( e^{bR_{0}}\left( N+bR\right) -N+bR\right)
+ab+2e^{bR_{0}}\right] e^{2bR}\right. =0.%
\end{array}%
\end{array}%
\right.
\end{array}
\label{FR1}
\end{equation}

Solving Eq. (\ref{FR1}), we obtain the parameters of $F(R)$ models as
\begin{equation}
\begin{array}{c}
\text{type}-I\text{:} \left\{
\begin{array}{c}
\lambda =\frac{R\left( n-1\right) e^{\xi R}}{n+\xi R}, \\
\\
\eta =-\frac{\left( 1+\xi R\right) }{R^{n-1}\left( n+\xi R\right) },%
\end{array}%
\right. , \\
\\
\text{type}-II\text{:} \left\{
\begin{array}{c}
\alpha =-\frac{1}{2R^{n-1}}, \\
\\
\beta =\frac{R^{n-1}}{2},%
\end{array}%
\right. , \\
\\
\text{type}-III\text{:} \left\{
\begin{array}{c}
m^{2}=R\left( \frac{n-1}{c_{2}}\right) ^{\frac{-1}{n}}, \\
\\
c_{1}=n\left( \frac{n-1}{c_{2}}\right) ^{\frac{1}{n}-1},%
\end{array}%
\right. , \\
\\
\text{type}-IV\text{:} \left\{
\begin{array}{c}
c=\frac{\left[ e^{bR}+e^{bR_{0}}\right] ^{2}\left( bR+1-e^{bR}\right) }{%
R^{N-1}\left[ N\left( e^{bR}+e^{bR_{0}}\right) +bR\right] \left(
e^{bR}-1\right) }, \\
\\
a=-\frac{R\left\{ \left( N-1\right) e^{2bR}+\left[ bR\left(
1+e^{bR_{0}}\right) +\left( N-1\right) \left( e^{bR_{0}}-1\right) \right]
e^{bR}\right\} e^{bR}}{\left( e^{bR}-1\right) ^{2}\left[ bR+N\left(
e^{bR}+e^{bR_{0}}\right) \right] },%
\end{array}%
\right. ,%
\end{array}%
\end{equation}

Equation (\ref{g(r)pure}) is similar to the solutions obtained for
Einstein-CIM-gravity's rainbow (Eq. (\ref{g(r)CIM})). Calculations show that
the Ricci scalar is $R=\frac{2d}{d-2}\Lambda $ and the Kretschmann scalar
has the following behaviors
\begin{eqnarray}
\lim_{r\longrightarrow 0}R_{\alpha \beta \gamma \delta }R^{\alpha \beta
\gamma \delta } &\longrightarrow &\infty ,  \nonumber \\
\lim_{r\longrightarrow \infty }R_{\alpha \beta \gamma \delta }R^{\alpha
\beta \gamma \delta } &\longrightarrow &\frac{8d}{\left( d-1\right) \left(
d-2\right) ^{2}}\Lambda ^{2},
\end{eqnarray}%
which confirm that, there is a curvature singularity at $r=0$. In addition,
we find that the asymptotical behavior of the mentioned spacetime is similar
to (a)dS Einstein-CIM-rainbow black holes. In other words, one can extract
charged solutions from the pure gravity if the $F(R)$ model and its
parameters are chosen suitably.

\begin{figure}[tbp]
$%
\begin{array}{c}
\epsfxsize=7cm \epsffile{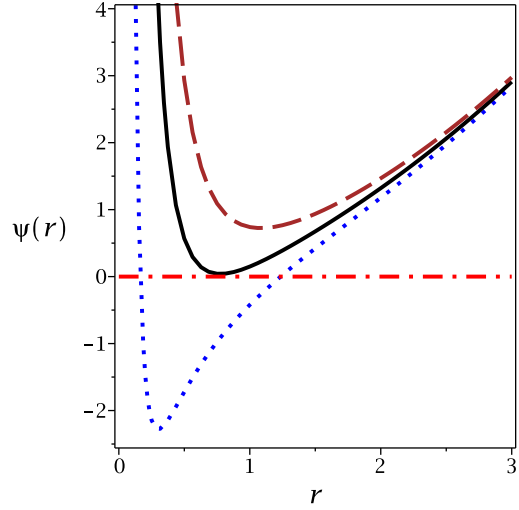}%
\end{array}
$%
\caption{$\protect\psi (r)$ versus $r$ for $m=2$, $d=4$, $f(E)=1.1$, $%
\Lambda =-1$, $g(E)=1.1$, $q=0.50$ (dotted line), $q=0.85$ (continuous line)
and $q=1.10$ (dashed line).}
\label{Figpure}
\end{figure}

For more investigation of these solutions, we plot $\psi (r)$ versus $r$.
Fig. \ref{Figpure} shows that, the mentioned singularity can be covered with
two horizons (Cauchy horizon and event horizon) or one horizon (extreme
black hole), and otherwise we encounter with a naked singularity.

\subsection{Thermodynamics}

In order to check the first law of thermodynamics for the obtained charged
black hole, first, we investigate the Hawking temperature of the pure $F(R)$
gravity's rainbow black holes on the outer horizon $r_{+}$ as
\begin{equation}
T_{+}=\frac{r_{+}^{d-2}\left[ \left( d-3\right) g^{2}(E)-\frac{2\Lambda
r_{+}^{2}}{d-2}\right] -2^{\left( d-4\right) /4}\left[ gf(E)g(E)\right]
^{d/2}}{4\pi f(E)g(E)r_{+}^{d-1}}.
\end{equation}

Using the method introduced in Ref. \cite{Cognola2005} for the obtained
solutions of pure $F(R)$ gravity's rainbow with $F(R)=F_{R}=0$, one may get
zero entropy ($S=\frac{A}{4}F_{R}=0$, $A$ is the horizon area). Therefore,
in order to resolve this problem and for obtain correct nonzero entropy, one
may consider the first law of thermodynamics as a fundamental principle. In
other words, we can use of $dS=\frac{1}{T}dM$. Using $\partial /\partial t$
as a Killing vector, one can extract the following relation for the finite
mass as
\begin{eqnarray}
M &=&\frac{(d-2)m}{16\pi f(E)g(E)^{d-3}} \\
&=&\frac{\left[ \left( d-2\right) g^{2}(E)-\frac{2\Lambda r_{+}^{2}}{\left(
d-1\right) }\right] r_{+}^{d-2}+2^{\left( d-4\right) /4}\left( d-2\right) %
\left[ gf(E)g(E)\right] ^{d/2}}{16\pi f(E)g(E)^{d-1}r_{+}}.
\end{eqnarray}

Hence, we can write
\begin{equation}
dM=\frac{r_{+}^{d-3}\left[ \left( \left( d-2\right) \left( d-3\right)
g^{2}(E)-2\Lambda r_{+}^{2}\right) r_{+}^{d-2}-2^{\left( d-4\right)
/4}\left( d-2\right) \left( gf(E)g(E)\right) ^{d/2}\right] }{16\pi
f(E)g^{d-1}(E)r_{+}^{d-1}}dr_{+},
\end{equation}%
so, we can obtain the entropy in the following form%
\begin{equation}
S=\int \frac{1}{T}dM=\frac{1}{4}\left( \frac{r_{+}}{g(E)}\right) ^{d-2},
\label{Pureentropy}
\end{equation}%
where it is the area law for the entropy. In other words, considering the $%
F(R)$ gravity's rainbow with $F_{R}=0$, one may use the so-called area law
instead of modified area law (Eq. (\ref{SFR})) \cite{Cognola2005}.

\subsection{DK Stability}

Here, we are going to discuss about the Dolgov-Kawasaki (DK) stability of
these solutions. DK stability in $F(R)$ gravity has been investigated in
literature \cite{Capozziello3,Cognola2005,AmendolaG,Bazeia,Rador,FaraoniA}.
It has been shown that there is no stable ground state for models of $F(R)$
gravity when $F(R)=0$ and also $F_{R}=dF(R)/dR\neq 0$ \cite{Schmidt}. It is
notable that, we find that $F(R)=F_{R}=0$, for the obtained solutions (in
order to obtain the charged black hole solution in $F(R)$ gravity with
constant Ricci scalar, Nojiri and Odintsov in Ref. \cite{NojiriOII} showed
that, $F(R)$ and $F_{R}$ must be zero). On the other hand, it has been shown
that $F_{RR}=d^{2}F(R)/dR^{2}$ is related to the effective mass of the
dynamical field of the Ricci scalar (see Refs. \cite{Dolgov,Sawicki}, for
more details). So, the positive effective mass is a requirement usually
referred to DK stability criterion which leads to stable dynamical field
\cite{FaraoniPRD,Bertolami}. In order to check this stability, we calculate
the second derivative of the $F(R)$ functions with respect to the Ricci
scalar for specific models in the following forms
\begin{equation}
F_{RR}=\left\{
\begin{array}{cc}
-\frac{\left( n-1\right) \left[ \xi R\left( n+\xi R\right) +n\right] }{%
R\left( n+\xi R\right) }, & \text{type}-I \\
&  \\
\frac{-\left( n-1\right) ^{2}}{R}, & \text{type}-II \\
&  \\
\frac{n-1}{R}, & \text{type}-III \\
&  \\
\begin{array}{c}
\frac{b^{2}R\left[ \left( n-1\right) e^{2bR}+\left( e^{bR_{0}}\left(
bR+n-1\right) +bR-n+1\right) e^{bR}-\left( n-1\right) e^{bR_{0}}\right] }{%
\left( e^{bR}-1\right) ^{2}\left[ ne^{bR}+\left( n+bR\right) e^{bR_{0}}%
\right] } \\
-\frac{\left( bR+1-e^{bR}\right) \left\{ n\left( 1-n\right) b^{3bR}+\tau
_{1}e^{2bR}-\tau _{2}e^{b\left( R+R_{0}\right) }+n\left( n-1\right)
e^{2bR_{0}}\right\} }{\left( e^{bR}+e^{bR_{0}}\right) \left( e^{bR}-1\right)
^{2}\left[ ne^{bR}+\left( n+bR\right) e^{bR_{0}}\right] R},%
\end{array}
& \text{type}-IV%
\end{array}%
\right. ,  \label{FRR}
\end{equation}%
where
\begin{eqnarray}
\tau _{1} &=&\left[ 2n\left( 1-n\right) +bR\left( bR-2n\right) \right]
e^{bR_{0}}+n\left( 1-n\right) +bR\left( bR-2n\right) , \\
\tau _{2} &=&\left[ n\left( n-1\right) +bR\left( bR+2n\right) \right]
e^{bR_{0}}+2n\left( 1-n\right) +bR\left( bR+2n\right) .
\end{eqnarray}%
\begin{figure}[tbp]
$%
\begin{array}{cc}
\epsfxsize=7cm \epsffile{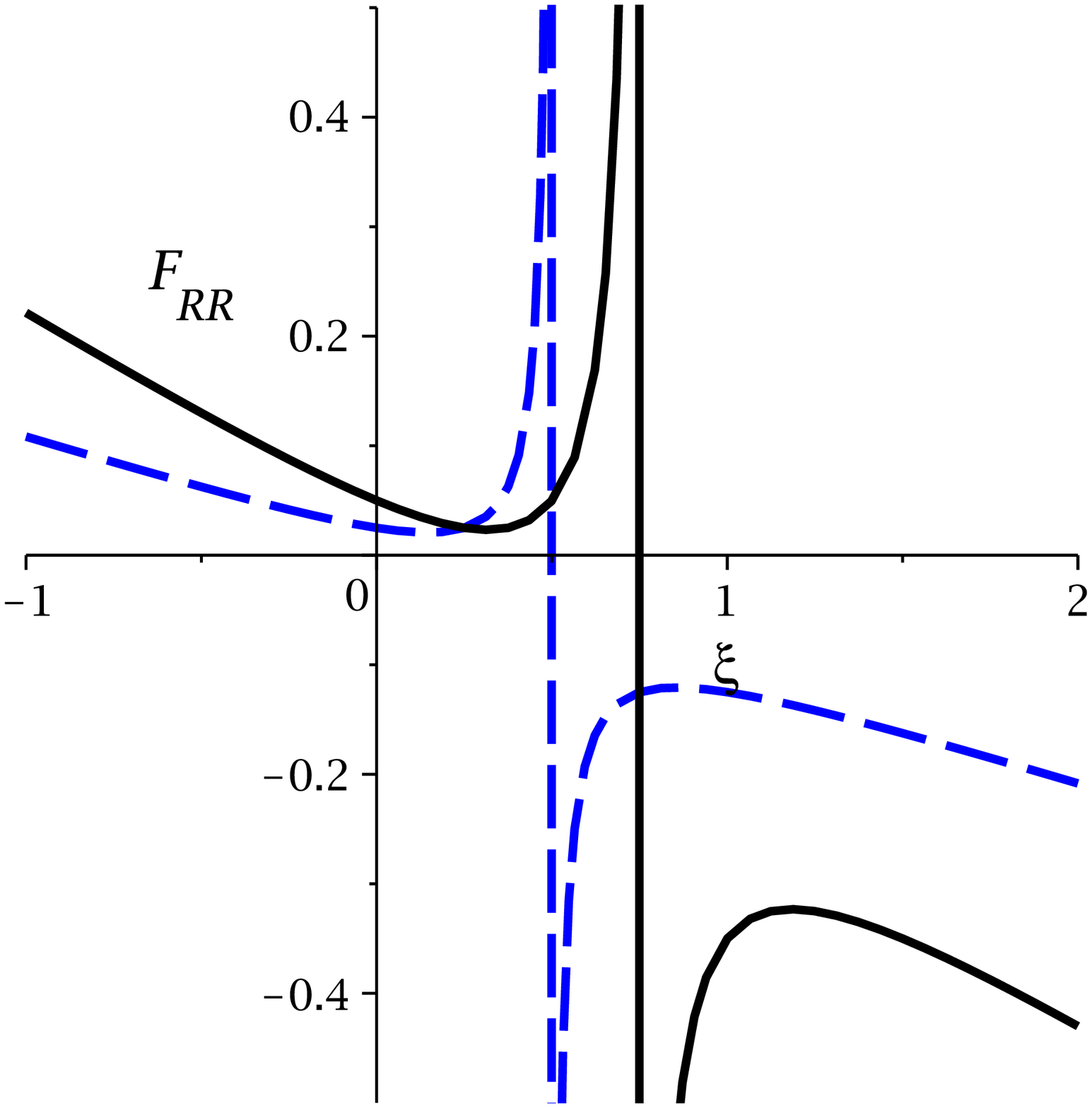} & \epsfxsize=7cm \epsffile{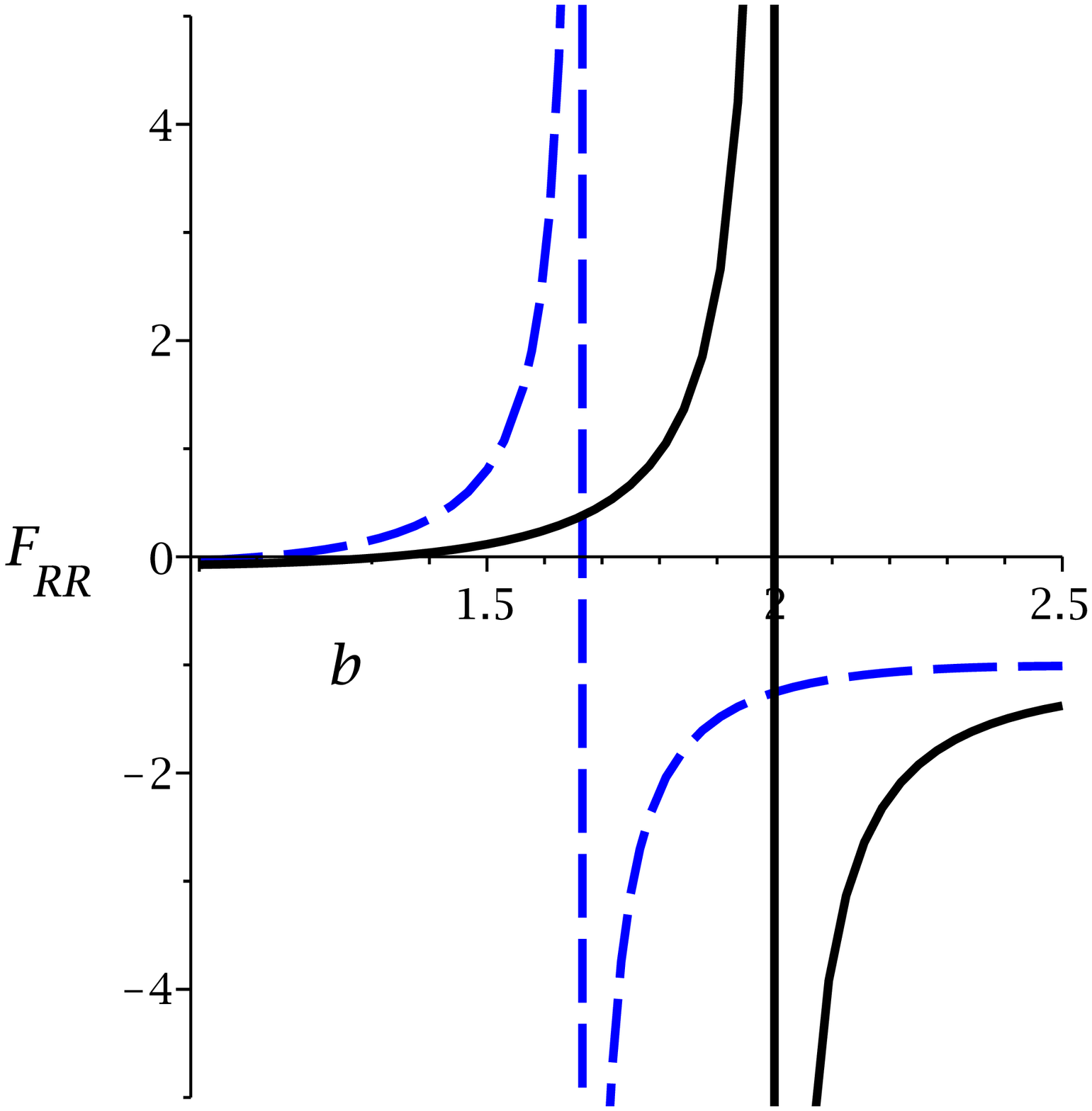}%
\end{array}
$%
\caption{type$-I$, left panel: $F_{RR}$ versus $\protect\xi $ for $\Lambda
=-1$ and $n=2$ (dashed line), $n=3$ (continues line). \newline
type$-IV$, right panel: $F_{RR}$ versus $b$ for $N=4$, $R_{0}=4$, $\Lambda
=-0.6$ (dashed line), $\Lambda =-0.5$ (continues line).}
\label{FigSTI}
\end{figure}

In order to have DK stability, $F_{RR}$ must be positive ($F_{RR}>0$). As
one can see, for type$-II$ and type$-III$ models, the sign of $F_{RR} $
depends on the sign of Ricci scalar. In other words, by considering $\Lambda
<0$ and $\Lambda >0$, the obtained solutions have DK stability for type$-II$
and type$-III$, respectively. On the other hand, for type$-I$ and type$-IV$,
the sign of $F_{RR}$ is not clear for arbitrary values of parameters. So, we
plot $F_{RR}$ in Fig. \ref{FigSTI}, to analyze the sign of $F_{RR}$. These
figures show that one may obtain stable solutions for special values of
parameters of models. In other words, we can set free parameters to obtain
stable models.

\section{Conclusions}

In this paper, we have considered higher dimensional Einstein-PMI and
Einstein-CIM theories in the presence of an energy dependent spacetime. We
obtained metric functions and discussed geometrical properties as well as
thermodynamical quantities. We showed that despite the contributions of the
gravity's rainbow in thermodynamical quantities, the first law of black
holes thermodynamics was valid. Next, we studied pure $F(R)$ gravity as well
as $F(R)$ gravity with CIM field in the presence of gravity's rainbow. We
pointed out that in case of pure $F(R)$ gravity's rainbow, for the suitable
choices of different parameters, one can obtain the electrical charge and
cosmological constant, simultaneously. In other words, we showed that, the
pure $F(R)$ gravity's rainbow was equivalent to the Einstein-CIM gravity's
rainbow. Also, in case of calculations of thermodynamical quantities for the
pure $F(R)$ gravity's rainbow, we pointed out that for these classes of
black holes, one should employ area law for obtaining entropy instead of the
usual modified approaches in case of $F_{R}=0$.

In addition, we conducted a study regarding the thermal stability of the
obtained solutions in this paper. We pointed out that variations of energy
functions in PMI class of the solutions lead to modifications in stability
conditions, phase transition points and thermal structure of the solutions
while the effects of such variations in $F(R)$ gravity were translation
like. We also observed that PMI parameter, $s$, and $F(R)$ gravity
parameter, $f_{R}$, were modifying factors in stability conditions and phase
transition points. Regarding the effect of dimensionality, we found that it
has a translation like behavior, too. An abnormal behavior was observed for
smaller divergency in $F(R)$ model.

Regarding the results of this paper, one may regard extended phase space and
$P-V$ criticality of the solutions. In addition, we can investigate
different viable models of $F(R)$ gravity to study the possible abnormal
behavior in thermal stability. Moreover, it is interesting to use a suitable
local transformation with appropriate boundary conditions to obtain the
so-called Nariai spacetime \cite{Nariai1,Nariai2} and discuss the
anti-evaporation process \cite{Noj1,Noj2,Noj3,Noj4}. We leave these issues
for future work.\newline

\begin{center}
\textbf{Conflict of Interests:}
\end{center}

The authors declare that there is no conflict of interests regarding the
publication of this paper.

\begin{acknowledgements}
We would like to thank the referee for constructive comments. We
thank Shiraz University Research Council. This work has been
supported financially by the Research Institute for Astronomy and
Astrophysics of Maragha, Iran.
\end{acknowledgements}

\end{document}